\documentclass[aps,prb,twocolumn,floats,epsfig,pdflatex]{revtex4-2}
\usepackage{amsmath}
\usepackage{amsfonts}
\usepackage{graphicx}
\usepackage{amssymb}
\usepackage{amsbsy}
\usepackage{comment}
\usepackage{color}
\usepackage[normalem]{ulem}

\newcommand{\beq}{\begin{equation}}
\newcommand{\eeq}{\end{equation}}
\newcommand{\bea}{\begin{eqnarray}}
\newcommand{\eea}{\end{eqnarray}}

\def\red{\textcolor{red}}

\begin{document}

\title{Floquet scars and prethermal fragmentation in a driven spin-one chain}
\author{Krishanu Ghosh$^1$, Diptiman Sen$^2$, and K. Sengupta$^1$}
\affiliation{$^1$School of Physical Sciences, Indian Association for the
Cultivation of Science, Kolkata 700032, India. \\
$^2$ Center for High Energy Physics, Indian Institute of Science, Bengaluru 560094, India.}
\date{\today}

\begin{abstract}

We study the periodic dynamics of a spin-one chain driven using a square-pulse protocol with amplitude $Q_0$ and frequency $\omega_D$. The Hamiltonian of the spin chain hosts a thermodynamically 
large number of $Z_2$-valued conserved quantities $W_{\ell}$ on the links $\ell$. This allows us to study the Floquet dynamics of this chain within a given sector with fixed values of $W_{\ell}$. For the sector with all $W_{\ell}=1$, we find signatures of quantum many-body scar states for $\hbar \omega_D \gg Q_0$; they lead to oscillatory dynamics and fidelity revival for specific initial states. Upon lowering $\omega_D$, we find an ergodic regime exhibiting fast thermalization consistent with the prediction of the (Floquet) eigenstate thermalization hypothesis. In addition, we identify special drive frequencies $\omega_n^{\ast}= Q_0/(2n \hbar)$ (where $n = 1, 2, 3, \cdots$) at which the Floquet Hamiltonian exhibits prethermal strong Hilbert space fragmentation (HSF) with the largest fragment being
ergodic; in contrast, a weak HSF is found at $\omega'_n= Q_0/[\hbar(2n+1)]$ (where $n = 0, 1, 2, \cdots$). We also study the sector with $W_{\ell} =\{\cdots 1,1,-1,1,1,-1 \cdots \}$ which shows strong HSF at $\omega_n^{\ast}$ but no fragmentation at $\omega'_n$. Our analysis indicates that the strong HSF in this sector harbors an integrable largest fragment. We provide numerical support for our analytical and perturbative results using exact-diagonalization (ED) studies on finite chains of length $L\le 24$. Our numerical results for entanglement entropy, fidelity, and correlation functions of the driven chain provide definitive signatures of prethermal strong HSF for both sectors.
\end{abstract}


\maketitle

\section{Introduction} 

The non-equilibrium dynamics of closed quantum systems have been intensely studied in recent years \cite{rev1,rev2,rev3,rev4,rev5,rev6,rev7,rev8,rev9,rev10,rev11,rev12,rev13,rev14,rev15,rev16,rev17}. Such studies have received support from experiments carried out with ultracold atoms in optical lattices \cite{rev13,rev14,exp1,exp2,exp3,exp4,exp5,exp6,exp7}. Recently, research in this area has focused on periodically driven systems. This is primarily due to the fact that such systems exhibit several phenomena that have no analogs in their aperiodically driven counterparts. Some examples of such phenomena include dynamical localization \cite{dynloc1,dynloc2,dynloc3,dynloc4,dynloc5,dynloc6,dynloc7} and transition \cite{dyntran1,dyntran2,dyntran3,dyntran4,dyntran5,dyntran6,dyntran7,dyntran8}, dynamical freezing \cite{dynfr1,dynfr2,dynfr3,dynfr4,dynfr5,dynfr6,dynfr7}, prethermal time crystalline behavior \cite{tcrev1,tcrev2,tcrev3,tcrev4,tcrev5,tc1,tc2,tc3,tc4,tc5}, generation of Floquet states with non-trivial topology \cite{topo1,topo2,topo3,topo4,topo5,topo6,topo7,topo8,topo9}, prethermal realization of quantum scars \cite{scar1,scar2,scar3,scar4,scar5,scar6,scar7,scar8}, Hilbert space fragmentation (HSF) \cite{sala2020,khemani2020,moudgalya2022,hsf1,hsf2,hsf3,hsf4,hsf5,hsf6,aditya2024,ganguli2025} and Meissner phases \cite{meissner1}. In addition, it has led to ways of reducing the heating in driven quantum systems \cite{berry1,rev18,tista1}. 

The time-evolution of a periodically driven system is controlled by its evolution operator $U(t,0)$. At stroboscopic times $t= mT$, where $T= 2\pi/\omega_D$ is the time period of the drive, $m \in Z$ is the number of drive cycles, and $\omega_D$ is the drive frequency, these operators are described by a Floquet Hamiltonian $H_F$ through the relation $U(mT,0) = \exp[-i H_F mT/\hbar]$, where $\hbar$ denotes Planck's constant. The exact computation of $H_F$ is usually difficult; for most system, one resorts to perturbative methods where the inverse of the drive amplitude or drive frequency is treated as the perturbation parameter \cite{rev10,rev12}. These methods are therefore useful in the large drive frequency or amplitude regime where the Floquet Hamiltonian resembles Hamiltonians with short-range interactions. In these regimes, the driven systems usually hosts a long prethermal timescale before eventual thermalization\cite{mori1,saito1,da1,da2,vk1,vk2,dc1}; the dynamics within this regime is typically controlled by the first term, $H_F^{(1)}$, of the perturbation series. It has been shown that $H_F^{(1)}$ may host additional emergent symmetries which control the dynamics in this prethermal regime; such emergent symmetries are key to understanding several properties of such driven systems \cite{rev17}. 

\begin{figure}
\rotatebox{0}{\includegraphics*[width=\linewidth]{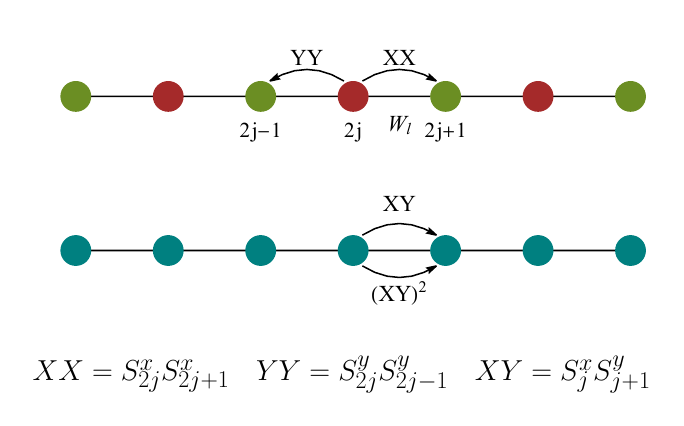}}
\caption{Schematic representation of the spin-one Kitaev chain. The top panel shows the representation in terms of $S^x_{2j} S^x_{2j+1}$ and $S^y_{2j} S^y_{2j-1}$ for even site $2j$ (shown as $XX$ and $YY$ respectively). For each of the links $\ell$ between any two sites, $W_{\ell}$, defined in Eq.\ \ref{cons1} of the main text, is conserved. The bottom panel shows an equivalent representation in terms of 
$XY \equiv S_{j}^x S_{j+1}^y$ on all sites of the chain; these two Hamiltonians are unitarily related as discussed in the text. An interaction term of the form $(XY)^2$ is also added; together they form the Hamiltonian $H$ studied in the text. \label{fig0}}
\end{figure}

In this work, we shall study the dynamics of a periodically driven spin chain. The Hamiltonian of the spin chain, shown schematically in the bottom panel of Fig.\ \ref{fig0}, is given by 
$H= H_0 + H_1$, where 
\begin{eqnarray} 
H_0 &=& J \sum_{j} S_j^x S_{j+1}^y, \quad H_1= Q \sum_j (S_j^x S_{j+1}^y)^2, \label{ham1}
\end{eqnarray} 
where $S_j^{\alpha}$ denotes spin-one ${\rm SU(2)}$ matrices on site $j$ of the chain \cite{sm1,sm2,dsen1}. We note at the outset that the Hamiltonian in Eq.~\ref{ham1} is equivalent to a spin-one Kitaev chain (shown schematically in the top panel of Fig.\ \ref{fig0}) studied in Refs.\ \onlinecite{sm1,sm2} via an unitary transformation of the spins $S_{2j}^x= S_{2j}^y$, $S_{2j}^y= S_{2j}^x$, and $S_{2j}^z=-S_{2j}^z$ on all the even sites of the chain. In what follows, we will drive $H_1$ according to a square pulse protocol given by
\begin{eqnarray} 
Q(t) &=& - ~Q_0 \quad {\rm for}\, \, ~ t\le T/2 \nonumber\\
&=& Q_0 \quad {\rm for}\,\, ~T/2 < t\le T, \label{sqprot} 
\end{eqnarray} 
in the large drive amplitude regime $Q_0 \gg J$. Our numerical study of this model will be facilitated by the fact that the model given by Eq.\ \ref{ham1} is known to host a large number of $Z_2$-valued conserved quantities $W_{\ell}$ which live on a link $\ell$ separating sites $j$ and $j+1$. These are given by
\begin{eqnarray} 
W_{\ell} &=& \Sigma_j^y \, \Sigma_{j+1}^x, \quad \Sigma_j^{\alpha}= e^{ i \pi S_j^{\alpha}}, \label{cons1} 
\end{eqnarray}
where $\alpha=x,y,z$. The existence of the $W_\ell$'s enables one to study the dynamics of the model within a given sector allowing for exact diagonalization (ED) of the spin chain of length $L\le 24$. In what follows we shall study the dynamics of the driven spin chain for two sectors; the first of these has $W_{\ell}=1$ for all links while the second has a periodic pattern given by $W_{\ell}= \{\cdots 1,1,-1,1,1,-1 \cdots\}$. Our numerical results will be supported by analytical, albeit perturbative, computations of the Floquet Hamiltonian in the large drive amplitude regime using Floquet perturbation theory (FPT) \cite{rev12}. 

The central results that we obtain from our studies are as follows. First we obtain an analytic expression for the first-order Floquet Hamiltonian $H_F^{(1)}$ in the strong drive amplitude regime. Our analysis indicates that $H_F^{(1)} \simeq H_0$ in the large drive frequency regime ($\hbar \omega_D \gg Q_0$). The system hosts quantum many-body scars in this regime which is expected to lead to oscillatory dynamical behavior for a class of initial states. Upon lowering the drive frequency, $H_F^{(1)}$ deviates from $H_0$ and the scar induced oscillatory dynamics is replaced by fast thermalization consistent with Floquet ETH.

Second, our analysis shows the existence of special drive frequencies $\omega_m^{\ast}= Q_0/(2n \hbar)$ ($n \in Z$) at which the driven system exhibits prethermal HSF \cite{hsf1,hsf2,hsf3,hsf4,hsf5,hsf6} for the sectors with
all $W_{\ell}=1$ and $W_{\ell} =\{\cdots 1,1,-1 \cdots \}$. We provide explicit analytic counting for such HSF by identifying and counting the number of states in the largest fragment of $H_F^{(1)}$ at these frequencies via transfer matrix methods. Our analysis shows that the largest fragment is ergodic in the sector with all $W_{\ell}=1$; in contrast, it is integrable for the $W_{\ell} =\{\cdots 1,1,-1 \cdots \}$ sector. 

Third, we find that for $\omega_D= \omega'_n= Q_0/[\hbar (2n+1)]$ ($n \in Z$), $H_F^{(1)}$ the driven chain exhibits weak prethermal fragmentation for $W_{\ell}=1$ sector. We numerically find that at these drive frequencies, the Hilbert space of $H_F^{(1)}$ supports dynamically disconnected fragments in the Fock basis. However, the number of such fragments does not grow exponentially with system size $L$; rather it grows only linearly with 
$L$. We note that the possibility of such different prethermal behaviors for different drive frequencies has been envisaged in Ref.\ \onlinecite{hsf5}; our model provides a concrete realization of this phenomenon. In contrast, the $W_{\ell}=\{\cdots 1,1,-1 \cdots \}$ sector 
does not exhibit any fragmentation for $\omega_D=\omega'_n$.

Fourth, we carry out exact numerical studies of quantum dynamics of the model using ED for both sectors. We compute the entanglement entropy $S$, the fidelity ${\mathcal F}(mT) =|\langle \psi_{\rm initial}|\psi(mT)\rangle|$ (where $m$ is the number of drive cycles), and the correlation function $C_{j,j+n} = \langle \psi(mT)|\hat O_j \hat O_{j+n}|\psi(mT) \rangle \equiv C_0 $, where $\hat O_j = (S_j^x S_{j+1}^y)^2$, for the driven chain as a function of $m$ at different drive frequencies. Our numerical results confirm the existence of quantum scars leading to oscillatory behaviors of ${\mathcal F}$ and $C_0$ at large drive frequencies for both sectors. Moreover, upon lowering $\omega_D$, these quantities exhibit rapid thermalization consistent with ETH. We also find distinct behaviors of the entropy, the fidelity, and the correlation function at $\omega_D=\omega_1^{\ast}$ and $\omega'_1$ as discussed below. 

For the sector with all $W_{\ell}=1$, the entanglement entropy of initial Fock states chosen randomly from the Hilbert space saturates to different Page values at $\omega_D=\omega_1^{\ast}$. This behavior is indicative of the presence of dynamically disconnected fragments in the Hilbert space. Moreover, for a chosen initial frozen state, $S$ remains pinned to its initial value $S_0$ for a large number of cycles; thus the deviation from $S$ from $S_0$ allows us to estimate the prethermal timescale up to which $H_F^{(1)}$ controls the dynamics. Our numerics indicates that the extent of this regime grows exponentially with the drive amplitude $Q_0$ which is consistent with the
results of Refs.\ \cite{hsf1}. For the $W_{\ell}=\{\cdots 1,1,-1\cdots\}$ sector, we find that for an initial state chosen from the largest fragment, the entanglement entropy exhibits oscillatory behavior between $S=0$ and $\ln 2$ which is consistent with the integrable nature of the largest fragment. We also find a distinct behavior of $S$ for the sector with all $W_{\ell}=1$ at $\omega_D= \omega'_1$. For this frequency, for several randomly chosen initial Fock states, $S/S_p$ saturates to a few distinct values at large times; this is consistent with the
presence of weak thermalization. 

The behavior of ${\mathcal F}$ also indicates the presence of such HSF at special frequencies. For the sector with all $W_{\ell}=1$ sector, for an initial state $|\psi_{\rm initial}\rangle$ chosen from the largest fragment, we find that the fluctuations of ${\mathcal F}$ at large times increase at $\omega_D=\omega_1^{\ast}$ indicating a much stronger finite-size effect; this arises due to the reduced Hilbert space dimension (HSD) of individual fragments. In contrast, for an initial state from the largest fragment in the $W_{\ell}=\{\cdots 1,1,-1\cdots\}$ sector, ${\mathcal F}$ shows an oscillatory behavior with perfect revivals for a long time indicating integrability of the fragment. Moreover, for an initial frozen state in this sector, ${\mathcal F}$ remains close to unity for a large number of cycles followed by a slow relaxation to its thermal value. The latter behavior occurs due o the presence of higher order terms in the Floquet Hamiltonian which do not host HSF. 

The characteristics of the driven system portrayed by $S$ and ${\mathcal F}$ receives further support from the behavior of $C_0$. In the $W_{\ell}=1$ sector, starting from the initial Fock state $|\psi_1\rangle= |\cdots x y
x y \cdots\rangle$, $C_0$ shows scar-induced oscillations for large drive frequencies, followed by fast thermalization as the frequency is lowered. In contrast, at $\omega_D= \omega_1^{\ast}$, $C_0$ remains pinned to its initial value for $m\le 200$ which is consistent with the behavior of the dynamics starting from a frozen state. A similar behavior is also found for the $W_{\ell}=\{\cdots 1,1,-1\cdots\}$ sector for the initial state $|\psi'_1\rangle=|\cdots x, y, x, x, y, x, \dots\rangle $. 

The plan of the rest of the paper is as follows. In Sec.\ \ref{model}, we chart out several properties of the Hamiltonian $H$ and provide an analytic computation of the HSDs of its different sectors using a transfer matrix method. This is followed by Sec.\ \ref{fptsec}, where we present an analytic computation of the first-order Floquet Hamiltonian $H_F^{(1)}$ for large $Q_0/J$ using Floquet perturbation theory (FPT) for both sectors; we also provide analytical and numerical estimates of the number of fragments and the HSD of the largest fragments of $H_F^{(1)}$ at $\omega_D= \omega_n^{\ast}$ and $\omega'_n$. Next, in Sec.\ \ref{numan}, we provide exact numerical results for $S$, ${\mathcal F}$ and $C_0$ using ED. Finally, we discuss our main results and conclude in Sec.\ \ref{diss}. 

\section{Model Hamiltonian and its sectors}
\label{model} 

In this section, we will chart out some basic properties of the spin chain. The analysis of the Hamiltonian and the presence of 
different dynamically disconnected sectors is discussed in Sec.\ \ref{modela} whereas the computation of dimension of these sectors, using 
a transfer matrix approach, is presented in Sec.\ \ref{modelb}. 

\subsection{ Properties of the Hamiltonian}
\label{modela} 

The Hamiltonian of the spin-one chain is given by Eq.\ \ref{ham1}. In what follows, we will work with the representation \cite{dsen1} 
\begin{eqnarray}
(S^{\alpha})_{\beta \gamma} = i \epsilon_{\alpha \beta \gamma}, \label{spinrep1}
\end{eqnarray}
where $\epsilon_{\alpha \beta \gamma}$ is the Levi-Civita antisymmetric tensor. We note that in the representation given by Eq.\ \ref{spinrep1}, the matrices $\Sigma_j^{\alpha}$ (Eq.\ \ref{cons1}) are diagonal and are given by
\begin{eqnarray} 
\Sigma_j^x &=& {\rm Diag}[1,-1,1], \quad \Sigma_j^y = {\rm Diag}[-1,1,-1], \nonumber\\
\Sigma_j^z &=& {\rm Diag}[-1,-1,1]. \label{mat1} 
\end{eqnarray}
It is easy to see that $\Sigma_j^x \Sigma_j^y \Sigma_j^z= I$, where $I$ is the $3 \times 3$ identity matrix for all $j$; moreover, they satisfy 
$[H, \Sigma_j^x \Sigma_{j+1}^y] =0$ implying that $W_{\ell} =\pm 1 $ are $Z_2$-valued conserved quantities.
This leads to an exponentially large number of 
dynamically disconnected sectors characterized by specific values of $W_{\ell}$ on each link \cite{sm1,sm2,dsen1}. In what follows we shall study the properties of $H$ in two sectors corresponding to all $W_{\ell}=1$ and $W_{\ell} = \{\cdots 1,1,-1\cdots \}$. The former sector has the largest HSD and hosts the ground state of the model while the latter hosts the first excited state. The dimensions of these sectors will be analyzed in the next subsection. 

The local basis vectors on any site $j$ which can be used to construct the Fock states of the model can be written as 
\begin{eqnarray} 
|x_j\rangle &=& \left(\begin{array}{c} 1 \\ 0 \\ 0 \end{array} \right) , \,\, |y_j\rangle =\left(\begin{array}{c} 0 \\ 1 \\ 0 \end{array} \right), \,\, |z_j\rangle = \left(\begin{array}{c} 0 \\ 0 \\ 1 \end{array} \right). \label{basis1} 
\end{eqnarray} 
The corresponding Fock states can be labeled as $|\cdots \alpha_{j-1} \beta_{j} \gamma_{j+1} \delta_{j+2} \cdots \rangle$, where $\alpha, \,\beta, \,\gamma,\,\delta$ take values $x$, $y$, or $z$. The advantage of this construction becomes apparent upon noticing that on any site $j$,
\begin{eqnarray} 
S_j^{\alpha} |\beta_j\rangle &=& - i \epsilon_{\alpha \beta \gamma} |\gamma_j\rangle, \nonumber\\ 
(S_j^{\alpha})^2 |\beta_j\rangle &=& (1- \delta_{\alpha \beta}) |\beta_j \rangle, \quad \Sigma_j^{\alpha} = 1-2 (S_j^{\alpha})^2 \label{sop1}
\end{eqnarray} 
so that the Fock states constructed out of these basis vectors are eigenstates of $H_1$. 

The construction of these basis vector also allows us to efficiently chart out the states in a given sector. For example, to have $W_{\ell}= \pm 1$ 
on any link $\ell$ separating the sites $j$ and $j+1$, we need the spin states on sites $j$ and $j+1$ to satisfy $\Sigma_j^y \Sigma_{j+1}^x |\alpha_j \beta_{j+1}\rangle=\pm 1$, where $\alpha$ and $\beta$ can take values $x,y,z$. A straightforward calculation shows that $W_{\ell}=1$ corresponds to the 
states with configurations 
\begin{eqnarray} 
& |x_j y_{j+1}\rangle, \, |y_j x_{j+1}\rangle, \, |z_j z_{j+1}\rangle, \, |z_j y_{j+1}\rangle,\,|x_j z_{j+1}\rangle, \label{plusst}
\end{eqnarray} 
while that for $W_{\ell}=-1$ they correspond to 
\begin{eqnarray} 
& |x_j x_{j+1}\rangle, \, |y_j y_{j+1}\rangle, \, |y_j z_{j+1}\rangle, \, |z_j x_{j+1}\rangle. \label{minusst}
\end{eqnarray} 
This allows us to construct the many-body Fock states in different sectors efficiently; any such Fock state $|p\rangle$ satisfies 
\begin{eqnarray}
H_1 |p\rangle = \epsilon_p|p\rangle, \label{eigen1} 
\end{eqnarray}
where $\epsilon_p= Q p$, and the value of the integer $p$ depends on the specific configuration of the basis vectors used to construct $|p\rangle$. We note that $|p\rangle$ 
represents degenerate states since multiple configurations involving different basis vectors at different sites can lead to the same $\epsilon_p$.

Finally, we note that the action of $H_0$ and $H_1$ on these states can be charted out in a straightforward manner. For states corresponding to $W_{\ell}=1$, we find that
\begin{eqnarray}
H_1 |x_j y_{j+1}\rangle &=& H_1 |z_j y_{j+1}\rangle = H_1 |x_j z_{j+1}\rangle = 0, \label{wl1h0ac}\\
H_1 |y_j x_{j+1}\rangle &=& Q |y_j x_{j+1}\rangle,\quad H_1 |z_j z_{j+1}\rangle = Q |z_j z_{j+1}\rangle, \nonumber
\end{eqnarray}
while for the states in the $W_{\ell}=-1$ sector, one has
\begin{eqnarray}
H_1 |x_j x_{j+1}\rangle &=& H_1 |y_j y_{j+1}\rangle = 0, \label{wl-1h0ac}\\
H_1 |z_j x_{j+1}\rangle &=& Q |z_j x_{j+1}\rangle,\quad H_1 |y_j z_{j+1}\rangle = Q |y_j z_{j+1}\rangle. \nonumber
\end{eqnarray}
The action of $H_0$ on these states can also be obtained in an analogous manner. For states in the $W_{\ell}=+1$ sector,
we find that
\begin{eqnarray}
H_0 |x_j y_{j+1}\rangle &=& H_0 |z_j y_{j+1}\rangle = H_0 |x_j z_{j+1}\rangle = 0, \label{wl1h1ac}\\
H_0 |y_j x_{j+1}\rangle &=& J |z_j z_{j+1}\rangle, \quad H_0 |z_j z_{j+1}\rangle = J |y_j x_{j+1}\rangle, \nonumber
\end{eqnarray}
while for states in $W_{\ell}=-1$ sector, one has
\begin{eqnarray}
H_0 |x_j x_{j+1}\rangle &=& H_0 |y_j y_{j+1}\rangle = 0, \\
H_0 |z_j x_{j+1}\rangle &=& -J |y_j x_{j+1}\rangle,~~ H_0 |y_j z_{j+1}\rangle = -J |z_j x_{j+1}\rangle. \nonumber
\end{eqnarray}
We will use these relations to derive the Floquet Hamiltonian of the driven spin chain in Sec.\ \ref{fptsec}. 

\subsection{Dimensions of different sectors}
\label{modelb}

In this section, we compute the HSD of the $W_{\ell}=1$ and $W_{\ell}=\{\cdots 1,1,-1 \cdots\}$ sectors of the spin model. 

We first consider the sector where $W_{\ell} = +1$ for all $\ell$. Guided by 
Eq.~\ref{plusst}, we find that the number of states in this sector for a 
system of length $L$ can be found by first defining a $8 \times 8$ transfer 
matrix $T^{+++}_{j,mn}$ at site $j$, where $m = (abc)$ and $n= (a'b'c')$
denote the row and column indices, and $(abc), ~(a'b'c')$ can take the eight 
possible values $xyx, ~xzy, ~xzz, ~yxy, ~yxz, ~zyx, ~zzy, ~zzz$. The
superscript $+++$ in $T_j^{+++}$ signifies that this transfer matrix covers 
four sites $(j,j+1,j+2,j+3)$ with states satisfying $W_\ell = W_{\ell+1} = W_{\ell+2} 
= +1$. 

The matrix element $T^{+++}_{j,mn}$ takes the value
1 if $m=(abc), ~n=(bcd)$, where $(abcd)$ is an allowed state at the four 
consecutive sites $(j,j+1,j+2,j+3)$ satisfying $W_\ell = W_{\ell+1} = W_{\ell+2} = 1$, 
while $T^{+++}_{j,mn} = 0$ in all other cases. To be explicit, we have
\beq T_j^{+++} ~=~ \left( \begin{array}{cccccccc}
0 & 0 & 0 & 1 & 1 & 0 & 0 & 0 \\
0 & 0 & 0 & 0 & 0 & 1 & 0 & 0 \\
0 & 0 & 0 & 0 & 0 & 0 & 1 & 1 \\
1 & 0 & 0 & 0 & 0 & 0 & 0 & 0 \\
0 & 1 & 1 & 0 & 0 & 0 & 0 & 0 \\
0 & 0 & 0 & 1 & 1 & 0 & 0 & 0 \\
0 & 0 & 0 & 0 & 0 & 1 & 0 & 0 \\
0 & 0 & 0 & 0 & 0 & 0 & 1 & 1 \end{array} \right). \label{T+++} \eeq
The eigenvalues of $T^{+++}_j$ are given by $\tau, ~-1/\tau$ and zero (appearing
six times), where $\tau = (\sqrt{5}+1)/2$ is the golden ratio. Since the 
largest eigenvalue of $T_j^{+++}$ is $\tau$, the number of states in a system 
of length $L$ grows as $\tau^L \simeq 1.618^L$ as $L \to \infty$. We note that the dependence of 
the HSD on $L$ is identical to that for PXP model describing experimentally realizable Rydberg chains 
\cite{exp1,exp2,exp3,exp4,fss1} hosting Floquet QMBS \cite{scar2,scar5}. 
Such scars have also been identified in the present model in Ref.\ \onlinecite{sm2}. 

Next, we consider the sector with $W_{\ell}= \{\cdots 1,1,-1\cdots \}$. To find the number of states in this sector, we
define three transfer matrices, denoted $T_j^{++-}$, $T_j^{+-+}$,
and $T_j^{-++}$ as described below. Then the number of
states in this sector for a system with $L$ sites will be given by 
$(\lambda_{max})^{L/3}$, where $\lambda_{max}$ is the largest eigenvalue of
the matrix $T_j^{++-} T_{j+1}^{+-+} T_{j+2}^{-++}$. 

The transfer matrix $T_{j,mn}^{++-}$ covers four sites $(j,j+1,j+2,
j+3)$, where $m =(abc)$ denotes states satisfying $W_\ell = W_{\ell+1} = +1$
and $n=(bcd)$ denotes states satisfying $W_{\ell+1} = +1, ~W_{\ell+2} = -1$. 
From Eqs.~\ref{plusst} and \ref{minusst}, we find that $T_j^{++-}$ must 
be a $8 \times 7$ matrix, with $m$ taking eight possible values
$xyx, ~xzy, ~xzz, ~yxy, ~yxz, ~zyx, ~zzy, ~zzz$, and $n$ taking seven 
possible values $xyy, ~xyz, ~xzx, ~yxx, ~zyy,~ zyz, ~zzx$. Using this, we find that
\beq T_j^{++-} ~=~ \left( \begin{array}{ccccccc}
0 & 0 & 0 & 1 & 0 & 0 & 0 \\
0 & 0 & 0 & 0 & 1 & 1 & 0 \\
0 & 0 & 0 & 0 & 0 & 0 & 1 \\
1 & 1 & 0 & 0 & 0 & 0 & 0 \\
0 & 0 & 1 & 0 & 0 & 0 & 0 \\
0 & 0 & 0 & 1 & 0 & 0 & 0 \\
0 & 0 & 0 & 0 & 1 & 1 & 0 \\
0 & 0 & 0 & 0 & 0 & 0 & 1 \end{array} \right). \label{T++-} \eeq
Next, $T_{j,mn}^{+-+}$ is a transfer matrix covering four sites 
$(j,j+1,j+2, j+3)$, where $m =(abc)$ denotes states satisfying 
$W_\ell = +1, ~W_{\ell+1} = -1$
and $n=(bcd)$ denotes states satisfying $W_{\ell+1} = -1, ~W_{\ell+2} = +1$. 
We find that $T_j^{+-+}$ must 
be a $7 \times 7$ matrix, with $m$ taking seven possible values $xyy, ~xyz, ~
xzx, ~yxx, ~zyy,~ zyz, ~zzx$, and $n$ taking seven possible values
$xxy, ~xxz, ~yyx, ~yzy, ~yzz, ~zxy, ~zxz$. From this observation, we find that
\beq T_j^{+-+} ~=~ \left( \begin{array}{ccccccc}
0 & 0 & 1 & 0 & 0 & 0 & 0 \\
0 & 0 & 0 & 1 & 1 & 0 & 0 \\
0 & 0 & 0 & 0 & 0 & 1 & 1 \\
1 & 1 & 0 & 0 & 0 & 0 & 0 \\
0 & 0 & 1 & 0 & 0 & 0 & 0 \\
0 & 0 & 0 & 1 & 1 & 0 & 0 \\
0 & 0 & 0 & 0 & 0 & 1 & 1 \end{array} \right). \label{T+-+} \eeq
Finally, $T_{j,mn}^{-++}$ is a transfer matrix covering four sites 
$(j,j+1,j+2, j+3)$, where $m =(abc)$ denotes states satisfying 
$W_\ell = -1, ~W_{\ell+1} = +1$
and $n=(bcd)$ denotes states satisfying $W_{\ell+1} = +1, ~W_{\ell+2} = +1$. 
We find that $T_j^{-++}$ must 
be a $7 \times 8$ matrix, with $m$ taking seven possible values $xxy, ~xxz, ~
yyx, ~yzy, ~yzz,~ zxy, ~zxz$, and $n$ taking eight possible values
$xyx, ~xzy, ~xzz, ~yxy, ~yxz, ~zyx, ~zzy, ~zzz$. This allows us to write
\beq T_j^{-++} ~=~ \left( \begin{array}{cccccccc}
1 & 0 & 0 & 0 & 0 & 0 & 0 & 0 \\
0 & 1 & 1 & 0 & 0 & 0 & 0 & 0 \\
0 & 0 & 0 & 1 & 1 & 0 & 0 & 0 \\
0 & 0 & 0 & 0 & 0 & 1 & 0 & 0 \\
0 & 0 & 0 & 0 & 0 & 0 & 1 & 1 \\
1 & 0 & 0 & 0 & 0 & 0 & 0 & 0 \\
0 & 1 & 1 & 0 & 0 & 0 & 0 & 0 \end{array} \right). \label{T-++} \eeq

Using Eqs.\ \ref{T++-}, \ref{T+-+} and \ref{T-++}, we find that 
the largest value of the magnitudes of the eigenvalues of
$T_j^{++-} T_{j+1}^{+-+} T_{j+2}^{-++}$
is about $3.7321$. We therefore conclude that the number of states in
this sector grows as $3.7321^{L/3} \simeq 1.5511^L$ as $L \to \infty$.
This agrees with the results in Ref.\ \onlinecite{sm1,sm2}. 

Overall, we find that the HSD of each sectors is drastically reduced due to the presence of conserved $W_{\ell}$. Since these sectors are dynamically disconnected, this allows us to study the dynamics of the spin chain in each sector using ED for $L\le 24$, a system size which is otherwise impossible 
to address via ED for a spin-1 chain. 

\section{Floquet perturbation theory}
\label{fptsec}

In this section we study the driven spin chain and obtain an analytic form of the first-order Floquet Hamiltonian for large drive amplitude. 
In what follows, we will drive $H_1$ using a square pulse protocol with a period $T=2 \pi/ \omega_D$, where $\omega_D$ is the drive frequency, described in Eq.\ \ref{sqprot}. We will restrict ourselves to the large drive amplitude regime for which $Q_0 \gg J$. In this case, the leading term in the evolution operator of the driven chain can be written as 
\begin{eqnarray} 
&& U(t,0) \equiv U_0(t) = e^{i Q_0 t \sum_j (S_j^x S_{j+1}^y)^2/\hbar} \quad {\rm for} ~\,\, t\le T/2 \nonumber\\
&& \quad = e^{ i Q_0 (T-t) \sum_j (S_j^x S_{j+1}^y)^2/\hbar} \quad {\rm for} ~\,\,T/2 < t\le T. \label{evol0} 
\end{eqnarray} 
At this order, we have $U_0(T,0)=I$ which yields $H_F^{(0)}=0$. 

The first-order correction to the evolution operator $U(T,0)$ is given by
\begin{widetext}
\begin{eqnarray}
U_1(T,0) &=& -~ \frac{i}{\hbar}~ \int_0^T dt U_0^{\dagger}(t,0) \,H_0 \, U_0(t,0) = U_1^{(1)}(T,0) + U_1^{(2)}(T,0), \nonumber\\
U_1^{(1)}(T,0) &=& -~ \frac{i}{\hbar}~ \int_0^{T/2} dt\, 
e^{-i Q_0 t \sum_j (S_j^x S_{j+1}^y)^2/\hbar} H_0 e^{i Q_0 t \sum_j (S_j^x S_{j+1}^y)^2/\hbar}, \nonumber\\
U_1^{(2)}(T,0) &=& -~ \frac{i}{\hbar} ~\int_{T/2}^T dt\, e^{ -i Q_0 (T-t) \sum_j (S_j^x S_{j+1}^y)^2/\hbar} H_0 e^{ i Q_0 (T-t) \sum_j (S_j^x S_{j+1}^y)^2/\hbar}. \label{u1eq1}
\end{eqnarray}
\end{widetext} 
We first compute the first term in Eq.\ \ref{u1eq1}. To this end, we resolve $U_0$ using the eigenstates $|p\rangle$ of $H_1$. These are the Fock states as can be seen from Eqs.\ \ref{wl1h0ac} and \ref{wl-1h0ac}. Using Eq.\ \ref{eigen1} with $\epsilon_p= p Q_0$, we can write, for $t \le T/2$, 
\begin{eqnarray} 
U_0(t,0) &=& \sum_p e^{i \epsilon_p t/\hbar} |p\rangle \langle p|. \label{mresolve1} 
\end{eqnarray} 
In terms of these eigenstates, we can write 
\begin{eqnarray} 
U_1^{(1)}(T,0) &=& -~ \frac{i T}{2\hbar}~ \sum_{p,q} \frac{\sin (\epsilon_p-\epsilon_q)T/(4\hbar)}{(\epsilon_p-\epsilon_q)T/(4\hbar)} \nonumber\\
&& \times ~e^{i(\epsilon_p-\epsilon_q)T/(4\hbar)} \langle p| H_0|q\rangle ~|p\rangle \langle q|. \label{u1eq2} 
\end{eqnarray} 
A similar result is easily obtained for $U_1^{(2)}$ leading to $U_1(T,0)= 2U_1^{(1)}(T,0)$. To make further progress, we need to analyze the matrix elements $\langle p|H_0|q\rangle$. The properties of these matrix elements depend on the sector being addressed. The analysis of the properties of these matrix elements is carried out for the $W_{\ell}=1$ sector in Sec.\ \ref{fpta} and the $W_{\ell}= \{\cdots 1,1,-1 \cdots\}$ sector in Sec.\ \ref{fptb}. 
\vspace*{.4cm}

\subsection{Analysis for the sector with all $W_{\ell}=1$}
\label{fpta} 

In this subsection, we will first analyze the matrix elements and derive an explicit form of the first-order Floquet Hamiltonian in Sec.\ \ref{flw1}. This will be followed by an explicit counting of the HSD of the Floquet Hamiltonian at the special drive frequencies in Sec.\ \ref{hsdwl1}.
\vspace*{.4cm}

\subsubsection{First-order Floquet Hamiltonian for the
sector with all $W_{\ell}=1$}
\label{flw1}

We begin by analyzing the matrix elements of $(H_0)_{pq} \equiv \langle p|H_0|q\rangle$. We find that in the $W_{\ell}=1$ sector, there are three classes of such terms. Using Eq.\ \ref{wl1h1ac}, we find that the only non-zero matrix elements correspond to 
\begin{widetext}
\begin{eqnarray}
T_1 &=&\langle \cdots x_{j-1} y_j x_{j+1} y_{j+2}\cdots |S_j^x S_{j+1}^y| \cdots x_{j-1} z_j z_{j+1} y_{j+2}\cdots \rangle, \nonumber\\
T_2 &=& \langle \cdots x_{j-1} y_j x_{j+1} z_{j+2}\cdots |S_j^x S_{j+1}^y| \cdots x_{j-1} z_j z_{j+1} z_{j+2}\cdots \rangle, \nonumber\\
T_3 &=& \langle \cdots z_{j-1} y_j x_{j+1} y_{j+2}\cdots |S_j^x S_{j+1}^y| \cdots z_{j-1} z_j z_{j+1} y_{j+2}\cdots \rangle, \nonumber\\
T_4 &=& \langle \cdots z_{j-1} y_j x_{j+1} z_{j+2}\cdots |S_j^x S_{j+1}^y| \cdots z_{j-1} z_j z_{j+1} z_{j+2}\cdots \rangle, \label{matel1}
\end{eqnarray}
\end{widetext}
and their Hermitian conjugates. From Eq.\ \ref{wl1h0ac}, one finds that $T_1$ corresponds to a process for which $\epsilon_p=\epsilon_q$ since $|\cdots x_{j-1} y_j x_{j+1} y_{j+2}\cdots \rangle$ and $| \cdots x_{j-1} z_j z_{j+1} y_{j+2}\cdots \rangle$ are eigenstates of $H_1$ with the
same eigenvalue. Similarly for $T_2$ and $T_3$, the states correspond to $\epsilon_p-\epsilon_q = Q_0$ and for $T_4$, 
they correspond to $\epsilon_p-\epsilon_q= 2 Q_0$. Moreover, inspecting the structure of the Fock states involved in these matrix elements one can define the operators
\begin{eqnarray} 
H_a &=& \sum_j P^x_{j-1} S_j^x S_{j+1}^y P_{j+2}^y,\nonumber\\
H_b &=& \sum_j (P_{j-1}^x P_{j+2}^z + P_{j-1}^z P_{j+2}^y) S_j^x S_{j+1}^y P_j^z P_{j+1}^z, \nonumber\\
H_c &=& \sum_j P^z_{j-1} S_j^x S_{j+1}^y P_j^z P_{j+1}^z P_{j+2}^z, \label{hamops} 
\end{eqnarray} 
where $P_j^{\alpha}= (1-(S_j^{\alpha})^2) $ is a projection operator which projects the local spin on site $j$ to $|\alpha\rangle$, with $|\alpha\rangle= |x\rangle, |y\rangle\, {\rm or} \, |z\rangle$. In terms of these operators, one can write $T_1= \langle p|H_a|q\rangle$, $T_2+T_3= \langle p| H_b|q\rangle$ and $T_4= \langle p|H_c|q\rangle$. 

Substituting Eqs.\ \ref{matel1} and \ref{hamops} in Eq.\ \ref{u1eq2}, we find 
\begin{eqnarray} 
U_1(T,0) &=& - ~\frac{i J T}{\hbar}~ \Big[H_a +\frac{\sin x}{x} (H_b e^{i x}+ {\rm H.c.}) \nonumber\\
&& ~~~~~~~~~~+ ~\frac{\sin 2x}{2x} (H_c e^{2ix} +{\rm H.c.}) \Big], \label{u1eq3}
\end{eqnarray}
where $x= Q_0 T/(4\hbar)$. Using the relation $H_F^{(1)}= i\hbar U_1(T,0)/T$, we finally obtain 
\begin{eqnarray}
H_F^{(1)} &=& J \Big[H_a +\frac{\sin x}{x} (H_b e^{i x}+ {\rm H.c.}) \nonumber\\
&& ~~~~+ ~\frac{\sin 2x}{2x} (H_c e^{2ix} +{\rm H.c.}) 
\Big]. \label{flham1} 
\end{eqnarray}

Eq.\ \ref{flham1} indicates that there are two classes of special drive frequencies. The first of these occur at $\omega_n^{\ast} = Q_0/(2 n \hbar)$, where $n \in Z$. At these frequencies, the first-order Floquet Hamiltonian is given by
$H_F^{(1)}= J H_a$. From Eq.\ \ref{hamops}, we find that $H_a$ involves projection operators which annihilates an exponentially large number 
of states in the Hilbert space of $H$. This is an indication that such a Floquet Hamiltonian may lead to a large number of frozen zero energy eigenstates and hence may support a Floquet realization of HSF \cite{hsf1}. 

The second class of such drive frequencies occurs at $\omega'_n= Q_0/[(2n+1) \hbar]$, where $H_F^{(1)} = H_a + H_b/[(n+1/2) \pi]$. As we will see in the next section, this leads to a weak 
HSF. We note these two frequencies will show distinct behaviors for small $n$; for larger $n$, the dynamics of the driven chain at these two frequencies will display close similarities due to the dominance of $H_a$ in $H_F^{(1)}$. 

\subsubsection{HSD at special drive frequencies for all $W_{\ell}=1$}
\label{hsdwl1}

To establish this intuition in a more concrete manner, we provide an explicit counting of the number of fragments corresponding to $H_a$. 
To this end, we resort to the transfer matrix formalism developed in Sec.\ \ref{modelb}. We note that $H_a$ 
can only produce transitions between the states 
$|\cdots x_{j-1} y_j x_{j+1} y_{j+2}\rangle$ and $|\cdots x_{j-1} z_j z_{j+1} y_{j+2}\cdots \rangle$ at any four sites $(j-1,j,j+1,j+2)$, 
since this is the only pair of states for which $\Delta \epsilon = (\epsilon_p-\epsilon_q) = 0$. To count the number of 
different fragments, we therefore remove states of the form $|\cdots x_{j-1} z_j z_{j+1} y_{j+2}\rangle$ at any four sites since 
such states lie in the same fragment as the 
state $|\cdots x_{j-1} y_j x_{j+1} y_{j+2}\rangle$ 
because these two states are connected to each other by $H_a$. 
The number of fragments is therefore given by another transfer matrix 
$S^{+++}_{j,mn}$, where $S^{+++}_{j,mn}$ is just obtained from $T^{+++}_{j,mn}$ (Eq.\ \ref{T+++}) 
by changing the matrix element $T^{+++}_{j,14}$ from $1$ to $0$. 
We thus have
\beq S_j^{+++} ~=~ \left( \begin{array}{cccccccc}
0 & 0 & 0 & 1 & 1 & 0 & 0 & 0 \\
0 & 0 & 0 & 0 & 0 & 1 & 0 & 0 \\
0 & 0 & 0 & 0 & 0 & 0 & 0 & 1 \\
1 & 0 & 0 & 0 & 0 & 0 & 0 & 0 \\
0 & 1 & 1 & 0 & 0 & 0 & 0 & 0 \\
0 & 0 & 0 & 1 & 1 & 0 & 0 & 0 \\
0 & 0 & 0 & 0 & 0 & 1 & 0 & 0 \\
0 & 0 & 0 & 0 & 0 & 0 & 1 & 1 \end{array} \right). \label{S+++} \eeq
We find that the largest value of the magnitudes of the eigenvalues of $S_j^{+++}$ is about $1.5499$. Hence the
number of fragments grows as $1.5499^L$ as $L \to \infty$.
\begin{table}[h!]
\centering
\renewcommand{\arraystretch}{1}
\begin{tabular}{|c|c|c|c|}
\hline
$L$ & $N_f$ & $N_{\rm fr}$ & ${\mathcal D}_L$ \\
\hline 
12 & 192 (192) & 118 (129) & 18 (18) \\
14 & 458 (462) & 281 (291) & 29 (29) \\
16 & 1107 (1108) & 673 (655) & 47 (47) \\
18 & 2667 (2663) & 1495 (1472) & 76 (76) \\
20 & 6403 (6398) & 3281 (3311) & 123 (123) \\
22 & 15370 (15370) & 7393 (7449) & 199 (199)\\
24 & 36920 (36922) & 16782 (16753) & 322 (322) \\
\hline
\end{tabular}
\caption{Comparison between exact numerics and analytical results (shown in brackets and rounded off to the nearest integer) for fragment number $N_f$, number of frozen fragments $N_{\rm fr}$, and the HSD of the largest fragment ${\mathcal D}_L$ for $H_F^{(1)} = H_a$ at $\hbar \omega_D=Q_0/(2n)$. All results are for the sector with all $W_{\ell}=1$. See text for details} \label{tabwl1}
\end{table} 

Next, we can find the size of the largest fragment as follows. After some 
inspection, we find that this is the fragment which contains the state
$|\cdots y_{j-1}x_j y_{j+1}x_{j+2} y_{j+3}x_{j+4} y_{j+5}x_{j+6} \cdots\rangle\equiv |\cdots yxyxyxyx \cdots \rangle$. 
The state of any four sites can change from $|xyxy \rangle$ 
to $| xzzy \rangle$. However, since $H_a$ does not allow the transitions $| zyxy \rangle
\leftrightarrow | zzzy \rangle$ and $| xyxz \rangle \leftrightarrow | xzzz \rangle$, a transition from
either $| xzzyxy \rangle$ or $| xyxzzy \rangle$ to $| xzzzzy \rangle$ is not allowed. Hence the state 
$| xyxyxy \rangle$ cannot change, in two steps, to $| xzzzzy \rangle$. Thus the fragment 
containing $| \cdots yxyxyxyx \cdots \rangle$ only contains states in which any two 
sites $yx$ can be replaced by $zz$ but two $zz$'s are not allowed to sit
next to each other. This resembles the PXP model in which there is a spin-1/2 
at each site and there is a constraint that two spin-1/2's cannot sit next to 
each other. 

To make this mapping from the sector with all $W_{\ell} = 1$ and with the constraint 
$\Delta \epsilon = 0$ to the PXP model more precise,
we replace the pair $yx$ by a spin-down and the pair $zz$ by a spin-up. If 
our model has $L$ sites, the PXP model has $L/2$ sites. The number of states
in the PXP model with $L/2$ sites is known to be $\tau^{L/2}$. Hence the
size of the largest fragment grows as $1.2720^L$. We thus
see that the ratio of the size of the largest fragment to the size of the
full sector is given by $\tau^{L/2} /\tau^L$ which goes to zero exponentially
as $0.7862^L$ as $L \to \infty$. Hence we have a strong prethermal HSF at the special
driving frequencies in the $W_{\ell}=1$ sector.

\begin{table}[htb]
\begin{center}
\begin{tabular}{|c|c|c|}
\hline
$L$ & $N_f$ & HSD of fragments\\
\hline
6 & 6 & {1, 1, 1, 1, 7, 7} \\
\hline
8 & 4 & {1, 15, 15, 16} \\
\hline
10 & 4 & {1, 31, 31, 60} \\
\hline
12 & 7 & {1, 1, 1, 1, 63, 63, 192}\\
\hline
14 & 5 & {1, 28, 127, 127, 560}\\
\hline
16 & 5 & {1, 160, 255, 255, 1536}\\
\hline
18 & 8 & {1, 1, 1, 1, 511, 511, 720, 4032}\\
\hline
20 & 6 &{1, 40, 1023, 1023, 2800, 10240}\\
\hline
22 & 6 & {1, 308, 2047, 2047, 9856, 25344}\\
\hline
24 & 9 & {1, 1, 1, 1, 1792, 4095, 4095, 32256, 61440}\\
\hline
26 & 7 & {1, 52, 8191, 8191, 8736, 99840, 146432}\\
\hline
28 & 7 & {1, 504, 16383, 16383, 37632, 295680, 344064}\\
\hline
\end{tabular}
\end{center}
\caption{Table showing the number of fragments $N_f$ and their HSD's for different $L$. 
We have taken the spin chain to have periodic boundary conditions; all results are for the
$W_{\ell} = 1$ sector, and $\omega_D= \omega'_1 = Q_0/\hbar$ where $H_F^{(1)}= H_a +2H_b/\pi$. See text for details.}
\label{table1} \end{table}

Finally, we can find the number of frozen fragments; these are
fragments which have only one state each. Hence these states
are zero energy eigenstates of the first-order Floquet Hamiltonian.
The number of frozen fragments can be found by simply removing the
two states $| xyxy \rangle$ and $| xzzy \rangle$ at any four sites since these are the 
only states between which transitions are allowed by the Hamiltonian. We can therefore define another
transfer matrix which is obtained from $T^{+++}_{j,mn}$ by setting
the matrix elements $(mn)= (14)$ and $(37)$ equal to zero. The
largest value of the magnitudes of the eigenvalues of this matrix
is found to be about $1.4997$. Hence the number of frozen fragments
grows as $1.4997^L$ as $L \to \infty$. These analytical results show excellent match with exact numerics as can be can be seen in Table \ \ref{tabwl1}. 

Eq.\ \ref{flham1} also indicates that for the special frequencies $\omega'_n = Q_0/((2n+1) \hbar)$, the first-order Floquet Hamiltonian contains $H_a$, $H_b$ and its Hermitian
conjugate but does not contain $H_c$. We will now shows that this
also leads to a HSF which is different from the case discussed
above where the first-order Floquet Hamiltonian only contains $H_a$.
This can be quantified as follows.

To find the number of fragments and the size of the largest
fragment in this case, we resort to a numerical counting for
the first few values of $L$ (chosen to be even) with periodic boundary conditions. These are shown in Table~\ref{table1}.
We see that as the system size $L$ increases,
the number of fragments grows only linearly (increasing by
$1$ whenever $L$ increases by $6$), while the size of
the largest fragment grows exponentially. Since we know that the
total number of states grows asymptotically as $1.618^L$, the
size of the largest fragment must also grow exponentially in the
same way if the number of fragments only grows linearly with $L$.

Finally, we can find the number of frozen fragments which have 
only one state each.
The number of such fragments can be found by removing the
six states having combinations $|xyxy \rangle$, $| xzzy
\rangle$, $| xyxz \rangle$, $| xzzz \rangle$, $| zyxy \rangle$ and $| zzzy \rangle$ at any four sites since these are the only states between which 
transitions are allowed by the Hamiltonian. We therefore define another
transfer matrix which is obtained from $T^{+++}_{j,mn}$ by setting
the matrix elements $(mn)= (14), ~(37), (15), ~(38), ~(64)$ 
and $(87)$ equal to zero. We find that this matrix has four
eigenvalues $\lambda_i$ equal to 1, and the other 
four eigenvalues are equal to zero. Hence, for a system with $L$ sites
and periodic boundary conditions, the number of frozen fragments
does not grow exponentially with $L$. We find that the number is
$4$ if $L$ is a multiple of 3 and is 1 if $L$ is not a 
multiple of 3 (see Table~\ref{table1}). This can be
understood as follows.
For all values of $L$, we find that 
one of the fragments consists of the state $|\cdots zzzzzz \cdots\rangle $.
When $L$ is a multiple of 3, we find three other frozen fragments
which are given by three period-three states; one of these is $|\cdots zyxzyx \cdots\rangle$ 
while the other two states are related to this state by translations.

Based on an inspection of systems with size 
up to $L=12$ and periodic boundary conditions, we have the following understanding of the growth of the number 
of fragments with $L$ given in Table~\ref{table1}. 
For this purpose, we use the
idea of a {\it root state}~\cite{aditya2024,ganguli2025}.
This is defined as one particular state defined in each fragment
such that all states in that fragment can be connected
to that state by repeated actions of the first-order Floquet Hamiltonian. The number of root states is equal to
the number of fragments. There are many ways
of choosing the root states; hence we have to 
specify them in a particular way as discussed below. 

We find that the root states
have one of the following forms. First, we have
the frozen fragments which have only one state each; as
discussed above, their number is four if $L$ is a multiple of 3 and unity if
$L$ is not a multiple of $3$. Next, there are two fragments with root states which have the states $|\cdots yxyx \cdots yx \cdots\rangle$
and $|\cdots xyxy \cdots xy\cdots \rangle$ from sites 1 to $L$; these fragments
have $2^{L/2} - 1$ states each. Finally, the other fragments have root states which have $|\cdots yxzyxzyxz \cdots zzzz \cdots\rangle$ from sites $1$ to $L$. Namely, these root states have some repeated strings of $yxz$'s followed at the end by 
one or more
$z'$s. The number of $yxz$ strings can range from $2, ~4, 
~6, ~\cdots$, up to
the largest even integer which is less than $L/3$. The
total number of fragments obtained in this way agrees with
the numbers $n_f$ given in Table~\ref{table1}.

We note that the higher order terms in the Floquet Hamiltonian do not support the two kinds of HSF discussed above (i.e., when only $H_a$ is present, and when
$H_a$, $H_b$ and its Hermitian conjugate are present).
However, as we will see in 
Sec.\ \ref{numan}, the dynamics of the driven chain will bear distinct signatures of the fragmentation over 
a long prethermal timescale in the regime of large drive amplitude. 

\subsection{Analysis for $W_{\ell}=\{\cdots 1,1,-1\cdots \}$ sector}
\label{fptb}
In this subsection, we analyze the matrix elements $\langle p|H_0|q\rangle$ for the $W_{\ell}=\{\cdots 1,1,-1\cdots\}$ sector. We derive the first-order Floquet Hamiltonian in Sec.\ \ref{flw-1}. This is followed by Sec.\ \ref{hsdw-1} where we find the HSD of the Floquet Hamiltonian at the special drive frequencies.

\subsubsection{First-order Floquet Hamiltonian for the $W_{\ell}=\{\cdots 1,1,-1\cdots\}$}
\label{flw-1}

To analyze the matrix elements in the $W_{\ell}=\{\cdots 1,1,-1 \cdots \}$ sector, we note that there are three possible combinations of the $Z_2$-valued conserved quantities,. Each of these connect distinct states via action of $H_0$. For example, for the arrangement $(1,1-1)$ in three consecutive links 
with the middle sites labeled by $j$ and $j+1$, we find the non-zero matrix elements to be 
\begin{widetext}
\begin{eqnarray}
T'_1 &=&\langle \cdots x_{j-1} y_j x_{j+1} x_{j+2}\cdots |S_j^x S_{j+1}^y| \cdots x_{j-1} z_j z_{j+1} x_{j+2}\cdots \rangle, \nonumber\\
T'_2 &=& \langle \cdots z_{j-1} y_j x_{j+1} x_{j+2}\cdots |S_j^x S_{j+1}^y| \cdots z_{j-1} z_j z_{j+1} x_{j+2}\cdots \rangle. \nonumber\\
\label{matel2}
\end{eqnarray}
The first of these processes corresponding to $T'_1$ has $\Delta \epsilon= \epsilon_p-\epsilon_q= Q_0$ while that corresponding to $T'_2$ has $\Delta \epsilon = 2Q_0$.
Next, for the arrangement $W_{\ell}= (1,-1,1)$, we find the matrix elements to be 
\begin{eqnarray}
T'_3 &=&\langle \cdots x_{j-1} y_j z_{j+1} y_{j+2}\cdots |S_j^x S_{j+1}^y| \cdots x_{j-1} z_j x_{j+1} y_{j+2}\cdots \rangle, \nonumber\\
T'_4 &=& \langle \cdots z_{j-1} y_j z_{j+1} y_{j+2}\cdots |S_j^x S_{j+1}^y| \cdots z_{j-1} z_j x_{j+1} y_{j+2}\cdots \rangle, \nonumber\\
T'_5 &=& \langle \cdots z_{j-1} y_j z_{j+1} z_{j+2}\cdots |S_j^x S_{j+1}^y| \cdots z_{j-1} z_j x_{j+1} z_{j+2}\cdots \rangle, \nonumber\\
T'_6 &=& \langle \cdots x_{j-1} y_j z_{j+1} z_{j+2}\cdots |S_j^x S_{j+1}^y| \cdots x_{j-1} z_j x_{j+1} z_{j+2}\cdots \rangle, \nonumber\\
\label{matel3}
\end{eqnarray}
which correspond to $\Delta \epsilon=0$ for $T'_3$ and $T'_5$ and $\Delta \epsilon= Q_0$ for $T'_4$ and $T'_6$. Finally, for the arrangement $W_{\ell}= (-1,1,1)$, one has
\begin{eqnarray}
T'_7 &=&\langle \cdots y_{j-1} y_j x_{j+1} y_{j+2}\cdots |S_j^x S_{j+1}^y| \cdots y_{j-1} z_j z_{j+1} y_{j+2}\cdots \rangle, \nonumber\\
T'_8 &=& \langle \cdots y_{j-1} y_j x_{j+1} z_{j+2}\cdots |S_j^x S_{j+1}^y| \cdots y_{j-1} z_j z_{j+1} z_{j+2}\cdots \rangle, \nonumber\\
\label{matel4}
\end{eqnarray}
\end{widetext}
which correspond to $\Delta \epsilon=Q_0$ and $\Delta \epsilon= 2Q_0$ respectively.

From these matrix elements it is easy to see, following the analysis of Sec.\ref{fpta}, that one may write down the operators 
\begin{eqnarray} 
H'_a &=& \sum_j (P^x_{j-1} P_{j+2}^y + P^z_{j-1} P_{j+2}^z) S_j^x S_{j+1}^y, \nonumber\\
H'_b &=& \sum_j \Big[(P_{j-1}^x P_{j+2}^x + P_{j-1}^y P_{j+2}^y) S_j^x S_{j+1}^y P_j^z P_{j+1}^z, \nonumber\\
&& + (P_{j-1}^z P_{j+2}^y + P_{j-1}^xP_{j+2}^z) S_j^x S_{j+1}^y P_j^z P_{j+1}^x \Big]\label{hamops2} \\
H'_c &=& \sum_j (P^z_{j-1} P_{j+2}^x + P_{j-1}^y P_{j+2}^z) S_j^x S_{j+1}^y P_j^z P_{j+1}^z. \nonumber
\end{eqnarray} 
In terms of these operators one can write $T'_3 +T'_5 = \langle p|H'_a|q\rangle$, 
$T'_1+T'_4+ T'_6+ T'_7 = \langle p| H'_b|q\rangle$ and $T'_2+T'_8= \langle p|H'_c|q\rangle$. 
Then an analysis exactly to the one carried out in Sec.\ \ref{fpta} yields
\begin{eqnarray}
H_F^{'(1)} &=& J~ \Big[H'_a +\frac{\sin x}{x} (H'_b e^{i x}+ {\rm H.c.}) \nonumber\\
&& ~~~~~+~ \frac{\sin 2x}{2x} ~(H'_c e^{2ix} +{\rm H.c.}) \Big]. \label{flham2} 
\end{eqnarray}
Once again, at the special frequency $\omega_n^{\ast}= Q_0/(2n\hbar)$, the first-order Floquet Hamiltonian is given by $J H'_a$. 
We therefore once again expect a realization of prethermal strong HSF at these frequencies. We will verify this expectation both analytically and numerically in the next subsection. 

\subsubsection{HSD at special drive frequencies for the sector $W_{\ell}=\{\cdots 1,1,-1 \cdots\}$}
\label{hsdw-1}

To put these ideas on a solid mathematical footing, we count the number of different fragments similar to that done for the $W_{\ell}=1$ sector. 
Here we note that the transitions
corresponding to $\Delta \epsilon= 0$ connect the states $ |\cdots x_{j-1}y_j z_{j+1} y_{j+2} \cdots \rangle \equiv |xyzy \rangle \leftrightarrow 
|\cdots x_{j-1}z_j x_{j+1} y_{j+2} \cdots\rangle \equiv |xzxy \rangle$, and the states $|zyzz \rangle \leftrightarrow |zzxz \rangle$ at any four 
sites $(j-1,j,j+1,j+2)$ for which $W_{\ell} = \{1,-1,1 \}$. Hence we remove states of the form 
$|xzxy \rangle$ and $|zzxz \rangle$ at such sites, since these states lie in the same 
fragment as the states $|xyzy \rangle$ and $|zyzz \rangle$ respectively. 
We therefore replace the transfer
matrix $T^{+-+}_{j,mn}$ by another transfer matrix $S^{+-+}_{j,mn}$, where 
$S^{+-+}_{j,mn}$ is obtained from $T^{+-+}_{j,mn}$ by changing the 
matrix elements $T^{+-+}_{j,36}$ and $T^{+-+}_{j,77}$ 
from $1$ to $0$. We thus have
\beq S_j^{+-+} ~=~ \left( \begin{array}{ccccccc}
0 & 0 & 1 & 0 & 0 & 0 & 0 \\
0 & 0 & 0 & 1 & 1 & 0 & 0 \\
0 & 0 & 0 & 0 & 0 & 0 & 1 \\
1 & 1 & 0 & 0 & 0 & 0 & 0 \\
0 & 0 & 1 & 0 & 0 & 0 & 0 \\
0 & 0 & 0 & 1 & 1 & 0 & 0 \\
0 & 0 & 0 & 0 & 0 & 1 & 0 \end{array} \right). \label{S+-+} \eeq
We then find that the largest value of the magnitudes of the
eigenvalues of $T_j^{++-} S_{j+1}^{+-+} T_{j+2}^{-++}$ is given by $3.1149$. 
Hence the
number of fragments grows as $3.1149^{L/3} \simeq 1.4604^L$ as $L
\to \infty$. We have checked that these results do not change if we remove
the states $|xyzy\rangle$ instead of $|xzxy \rangle$, and $|zyzz \rangle$ instead of $|zzxz \rangle$.

Next, we compute the number of frozen fragments;
these are fragments with only one Fock state. To do this, we
remove the states $|xyzy \rangle, ~|xzxy \rangle, ~|zyzz
\rangle$ and $| zzxz \rangle$
at any four sites $(j-1,j,j+1,j+2)$ for which $W_\ell = {1, -1, 1}$ since these states are connected to each other
by the Floquet Hamiltonian.
We therefore replace the transfer
matrix $T^{+-+}_{j,mn}$ by another transfer matrix $R^{+-+}_{j,mn}$, where 
$R^{+-+}_{j,mn}$ is obtained from $T^{+-+}_{j,mn}$ by changing the 
matrix elements $T^{+-+}_{j,24}$, $T^{+-+}_{j,36}$,
$T^{+-+}_{j,65}$ and $T^{+-+}_{j,77}$ 
from $1$ to $0$. We thus have
\beq R_j^{+-+} ~=~ \left( \begin{array}{ccccccc}
0 & 0 & 1 & 0 & 0 & 0 & 0 \\
0 & 0 & 0 & 0 & 1 & 0 & 0 \\
0 & 0 & 0 & 0 & 0 & 0 & 1 \\
1 & 1 & 0 & 0 & 0 & 0 & 0 \\
0 & 0 & 1 & 0 & 0 & 0 & 0 \\
0 & 0 & 0 & 1 & 0 & 0 & 0 \\
0 & 0 & 0 & 0 & 0 & 1 & 0 \end{array} \right). \label{R+-+} \eeq
We find that the largest value of the magnitudes of the
eigenvalues of $T_j^{++-} R_{j+1}^{+-+} T_{j+2}^{-++}$ is given by $2.5616$. 
Hence the
number of fragments grows as $2.5616^{L/3} \simeq 1.3683^L$ as $L
\to \infty$.

Finally, we find the largest fragment as follows. We first note that
the state at the site $j$ which lies between the bonds with $W_{\ell} = W_{\ell+1} 
= +1$ does not change at all under transitions for which $\epsilon= 0$.
Hence we only have to consider the sites which lie at the ends of the
bonds with $W_{\ell+2}=-1$. After some inspection, we find that the largest
fragment is the one which contains the states $|zyzz \rangle$ and $|zzxz \rangle$ at
sites $(j-1,j,j+1,j+2)$ which have $W_{\ell} = 1, ~W_{\ell+1}=-1, ~W_{\ell+2} = 1$.
Since there are two possible states for every sequence of three such bonds, 
the size of this fragment is $2^{L/3} = 1.2599^L$. The ratio of the
size of the largest fragment to the size of the full sector is given by
$1.2599^L / 1.5511^L = 0.8123^L$ which goes to zero exponentially as $L \to
\infty$. This proves the presence of prethermal HSF at the special frequencies. 
The analytical results obtained here matches those obtained from exact numerics quite well as can be seen from Table\ \ref{tabwl-1}.

\begin{table}[h!]
\centering
\renewcommand{\arraystretch}{1.3}
\begin{tabular}{|c|c|c|c|}
\hline
$L$ & $N_f$ & $N_{\rm fr}$ & ${\mathcal D}_L$ \\
\hline
6 & 11 (10) & 10 (7) & 4 (4) \\
12 & 95 (94) & 50 (43) & 16 (16) \\
18 & 914 (913) & 298 (283) & 64 (64) \\
24 & 8863 (8862) & 1890 (1854) & 256 (256) \\
\hline
\end{tabular}
\caption{Comparison between exact numerics and 
analytical results (shown in brackets and rounded off to 
the nearest integer) for the number of fragments $N_f$, number of frozen
fragments $N_{\rm fr}$, and the HSD of the largest fragment for different $L$, for the sector with $W_{\ell}=\{\cdots 1,1,-1\cdots\}$ and $\hbar \omega_D =
Q_0 /(2 n)$. See text for details.} \label{tabwl-1}
\end{table}

Since the largest fragment only contains Fock states with the sequences $| zyzz \rangle$ and $| zzxz \rangle$, it possesses an integrable structure. 
To see this, we note that the dynamics of such states can be described by a Hamiltonian of the form 
\begin{eqnarray}
H_{\rm eff} ~=~ J~ \sum_k ~\tau_k^x, \label{oscham} 
\end{eqnarray}
where $k$ takes $L/3$ possible
values, and $\tau_k^z = +1$ and $-1$ denote the states $| zyzz \rangle$ and $| zzxz \rangle$)
respectively. This constitutes a two-level (or, equivalently, decoupled spin-1/2) system for a given $k$. The
dynamics in this sector is therefore particularly simple and mimics that of an integrable model. 

For $\hbar \omega_D= Q_0/(2n+1)$, where $n \in Z$, $H_F= H'_a + H'_b \pi/(n+1/2)$; we numerically find that in contrast to the $W_{\ell}=1$ sector, the Hilbert space of $H_F$ has a single fragment. Thus these frequencies do not lead to either strong or weak fragmentation (i.e., 
exponential or power-law growth of the number of fragments with $L$) for the $W_{\ell} =\{\cdots 1,1,-1\cdots\}$ sector. 

In the next section, we shall study the dynamics in both these sectors numerically using ED
with the aim of finding signatures of strong HSF at the special drive frequencies. 

\section{Numerical results}
\label{numan} 

In this section, we provide numerical support for our analytical results derived in Sec.\ \ref{fptsec}. We study the properties of the half-chain 
entanglement entropy of the driven chain in Sec.\ \ref{entang}. This is followed in Sec.\ \ref{fidel} by a study of the fidelity of the driven state, starting from simple initial Fock states. Finally, we study the dynamics of a correlation function of the driven model in Sec.\ \ref{corr}.

\subsection{Entanglement entropy} 
\label{entang} 

\begin{figure}
\rotatebox{0}{\includegraphics*[width=0.49 \linewidth]{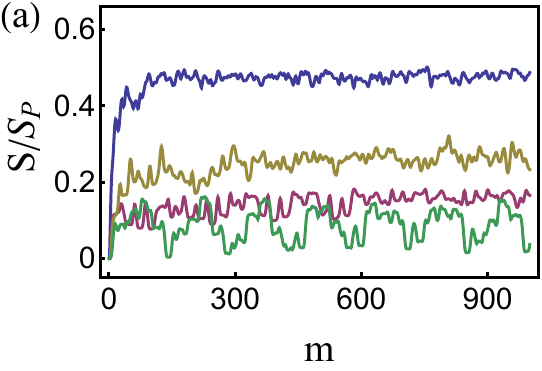}}
\rotatebox{0}{\includegraphics*[width=0.49 \linewidth]{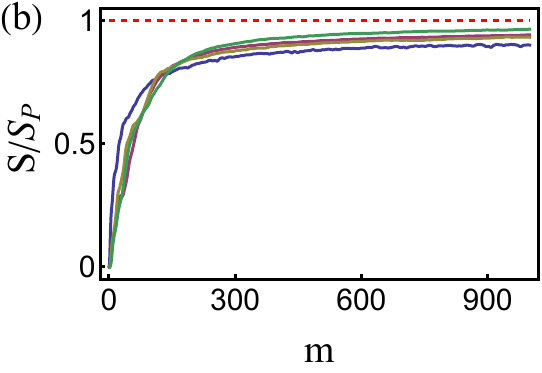}}
\caption{(a) Plot of the half-chain entanglement $S_{L/2}(mT)/S_p \equiv S/S_p$ as a function of
$m$ at the special frequency $\omega_D= \omega_1^{\ast}= Q_0/(2\hbar)$ and for several different representative initial Fock states. 
The plot indicates that the saturation value of $S/S_p <1 $ at large $m$; moreover, it depends on the 
chosen initial state. (b) In contrast, for $\omega_D= 1.2 \omega_1^{\ast}$, an analogous plot for the same initial states leads to $S/S_p \to 1$. For both plots, $L=24$, 
the initial states are chosen from the $W_{\ell}=1$ sector, $Q_0/J=100$, and all energies are in units of $J$.} \label{fig1}
\end{figure} 

In this subsection we study the half-chain entanglement entropy $S_{L/2} \equiv S$ of the driven chain starting from several initial Fock states. To compute $S$, we follow the procedure outlined in Refs.\ \onlinecite{scar1,scar4}. We start from an initial Fock state and evolve it up to $m$ drive cycles to obtain the many-body state $|\psi(mT)\rangle = U(mT,0)|\psi(0)\rangle$. Here $U(mT,0)$ is the evolution operator of the system 
and $|\psi(0)\rangle$ is the initial state. To compute $U(mT,0)$ for a square pulse protocol, we first numerically diagonalize the Hamiltonian $H_{\pm}= H[\pm Q_0]$ and find the corresponding eigenvalues $\epsilon_{\mu}^{\pm}$ and eigenstates $|\mu_{\pm}\rangle$ via
\begin{eqnarray} 
H_{\pm} |\mu_{\pm}\rangle= \epsilon_{\mu}^{\pm} |\mu_{\pm} \rangle. \label{eigen2} 
\end{eqnarray} 
Using these eigenstates and eigenvalues, we can express the evolution operator as 
\begin{eqnarray} 
U(mT,0) &=& \sum_{\mu \nu} c_{\mu \nu}^{\pm} \, e^{-i (\epsilon_{\mu}^+ + \epsilon_{\nu}^-) mT/(2\hbar)} \, |\mu_+\rangle \langle \nu_-|, \label{unmerical}
\end{eqnarray} 
where $c_{\mu \nu}^{\pm} = \langle \mu_+|\nu_-\rangle$. The density matrix corresponding to the state $|\psi(mT)\rangle$ can then be obtained as $\rho(mT)= |\psi(mT)\rangle \langle \psi(mT)|$. The reduced density matrix, computed from $\rho(mT)$ by numerically tracing out states which have weight on the subsystem $B$ of length $L/2$ of the chain, is given by \cite{scar1,scar4} 
\begin{eqnarray} 
\rho_{\rm red}(mT) &=& {\rm Tr}_B [\rho(mT)]. \label{redden1}
\end{eqnarray} 
The entanglement is obtained from the eigenvalues $\lambda_i(mT) $ of the reduced density matrix. These eigenvalues are obtained by numerical diagonalization of $\rho_{\rm red}(mT)$ and the 
von Neumann entanglement entropy $S$ is then obtained using 
\begin{eqnarray} 
S(mT) &=& \sum_i \lambda_i(mT) \ln \lambda_i(mT). \label{vn1}
\end{eqnarray} 
It is well-known that for an ergodic system, $S(mT)$ reaches its Page value, $S_p$, for $m\gg 1$ \cite{page1,dyntran5,hsf1}. 

\begin{figure}
\rotatebox{0}{\includegraphics*[width=0.49 \linewidth]{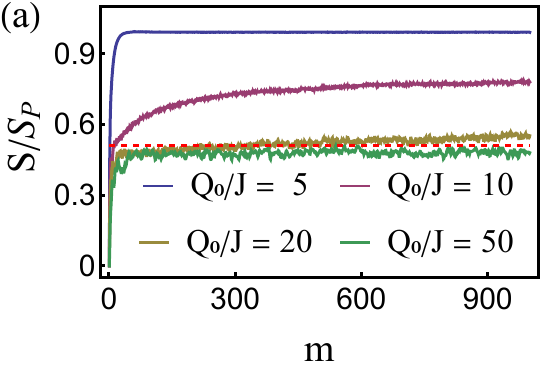}}
\rotatebox{0}{\includegraphics*[width=0.49 \linewidth]{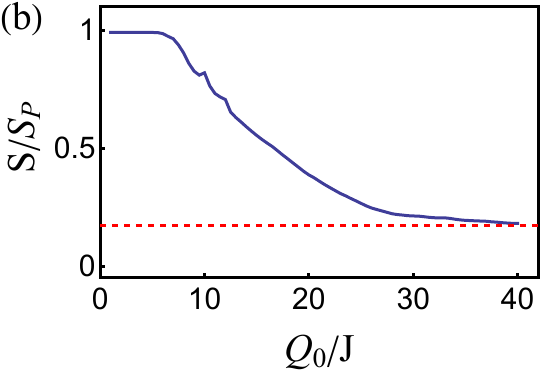}}
\caption{(a) Plot of $S_{L/2}(mT)/S_p \equiv S/S_p$ as a function of
$n$ at the special frequency $\omega_D= \omega_1^{\ast}= Q_0/(2\hbar)$ and for several representative values of 
$Q_0/J$. The initial state is chosen 
to be a Fock state lying in the largest fragment of $H_a$; the corresponding Page value of the fragmented sector, $S_p^f$, is indicated by the red-dashed line. (b) Plot of $S_{\rm av}/S_p$ as a function of $Q_0/J$ showing a gradual crossover from ergodic to fragmented dynamics. For both plots, $L=24$, the initial states are chosen from the $W_{\ell}=1$ sector, $\omega_D=\omega_1^{\ast}$, and all energies are scaled in units of $J$. } \label{fig2}
\end{figure} 

We first consider the dynamics of $S(mT)$ for the $W_{\ell}=1$ sector. The plot of $S(mT)$ as a function of $m$ is shown in Figs.\ \ref{fig1}(a) and (b). Fig.\ \ref{fig1}(a) shows that at the special drive frequencies $\omega_D= \omega_1^{\ast}$, $S(mT)/S_p < 1$ for all initial states; moreover the value to which $S(mT)/S_p$ saturates depends strongly on the chosen initial state. This indicates a retainment of the memory of the initial state and thus constitutes a clear violation of Floquet ETH which predicts $S/S_p \to 1$ for $m \to \infty$. Indeed the latter behavior is clearly seen when one moves way from the special frequencies; for any initial state, $S(mT)/S_p$ approaches unity for large $m$ as expected from Floquet ETH. 

The deviation of $S(mT)/S_p$ from unity at large $m$ and its dependence on the initial state is consistent with fragmentation of the Hilbert space at large drive amplitude where $H_a$ dominates the dynamics. In the presence of such a fragmentation, the entanglement of a driven state saturates to the Page value of the sector to which it belongs, and different choices of the initial state lead to different values of the entanglement at large prethermal times \cite{hsf1}. 

\begin{figure}
\rotatebox{0}{\includegraphics*[width=0.49 \linewidth]{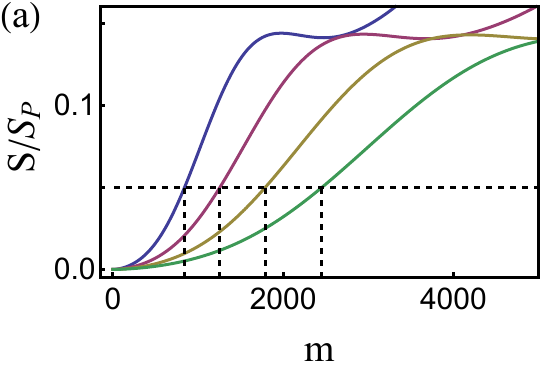}}
\rotatebox{0}{\includegraphics*[width=0.49 \linewidth]{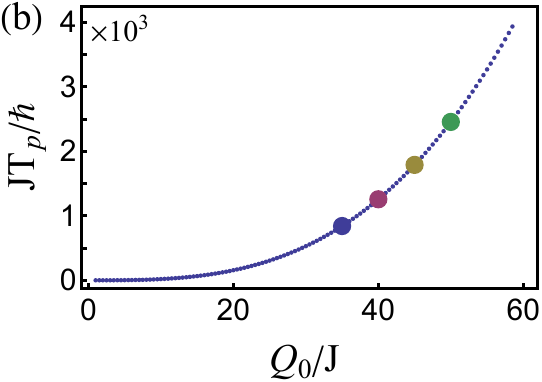}}
\caption{(a) Plot of $S_{L/2}(mT)/S_p \equiv S/S_p$ as a function of
$m$ at the special frequency $\omega_D= \omega_1^{\ast}= Q_0/(2\hbar)$ and $Q_0/J=35$ (blue), $40$ (magenta), $45$ (yellow), and $50$ (green) for an initial frozen state $|\psi(0)\rangle= |\cdots zzz\cdots\rangle$ lying in the $W_{\ell}=1$ sector. The dynamics of $S/S_p$ is a consequence of higher order terms in the Floquet Hamiltonian. The prethermal timescale $T_p= m_p T$ is estimated from the number of drive cycles $m=m_p$ for which $S(mT)/S_p \simeq 0.05$. (b) Plot of $JT_p/\hbar$ as a function of $Q_0$ keeping $\omega_D$ fixed at $\omega_1^{\ast}$; the data corresponding to panel (a) is shown by circles with same color code as in (a). For both
plots, $L=24$ and all energies are scaled in units of $J$. See text for details. } \label{fig3}
\end{figure} 

Next, we study the behavior of $S(mT)$ as a function of $m$ at the special frequency $\omega_D= \omega_1^{\ast}$ for several representative values of $Q_0$. For this we have chosen the initial state to be $|\psi_1 \rangle= |\cdots x y x y \cdots \rangle$ which belongs to the largest sector of $H_a$. The plot showing the behavior $S(mT)/S_p$ for different $Q_0$ is shown in Fig.\ \ref{fig2}(a). We find that at low $Q_0$, the driven chain exhibits ergodicity and $S/S_p \to 1$ at large $m$. In contrast,
for large $Q_0/J$, $S(mT)$ saturates to the Page value of the largest sector of $H_a$ given by $S_p^f$ shown by the red dotted line. In Fig.\ \ref{fig2}(b), we plot the average entanglement $S_{\rm av}/ S_p$ as a function of $Q_0$ at $\omega_D=\omega_1^{\ast}$, where $S_{\rm av}$ is given by
\begin{eqnarray} 
S_{\rm av} &=& \frac{1}{51} \sum_{n=325}^{375} S(mT). \label{avS}
\end{eqnarray} 
We find that $S_{av}$ interpolates between $S_p$ at lower $Q_0/J$ to $S_p^F$ at higher $Q_0/J$, indicating a gradual 
crossover to a prethermal state exhibiting Floquet HSF. 

To estimate this prethermal timescale over which the fragmentation occurs, we compute $S(mT)$ starting from a frozen product state of $H_a$ given by $|\psi_2 \rangle= |\cdots zzzz\cdots\rangle$. We note that if the dynamics is entirely controlled by $H_F^{(1)}$, $S(mT)=0$ at the special frequency. Thus the deviation of $S(mT)$ from its zero value provides an estimate of the prethermal timescale beyond which corrections to $H_F^{(1)}$ becomes important. A plot of $S(mT)/S_p$ for $\omega_D=\omega_1^{\ast}$ for several representative values of $Q_0/J$ is shown in Fig. \ref{fig3}(a). This allows us to define the number of cycles $m=m_p$ for which $S(mT)/S_p \simeq 0.05$. We find that $m_p$ increases with $Q_0/J$ and provides an estimate of the prethermal timescale $T_p= m_p T$. A plot of the variation of $T_p$ as a function of $Q_0/J$ is shown in Fig.\ \ref{fig3}(b). We find that $T_p$ grows exponentially with $Q_0$ which indicates that the prethermal timescale is exponentially large at large drive amplitudes. This conforms with the standard expectation for the dependence of the width of the prethermal regime on the drive amplitude~\cite{mori1,saito1,hsf1}. 

\begin{figure}
\rotatebox{0}{\includegraphics*[width=0.49 \linewidth]{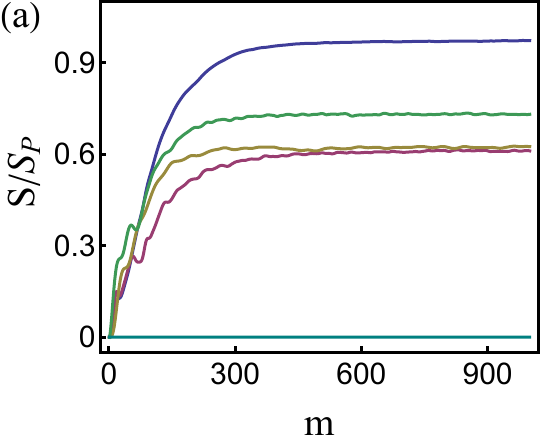}}
\rotatebox{0}{\includegraphics*[width=0.49 \linewidth]{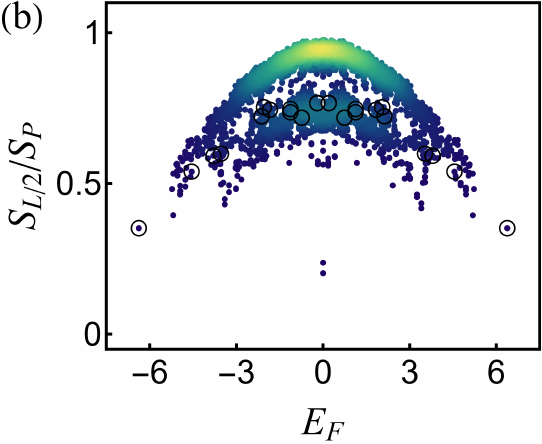}}
\caption{(a) Plot of $S_{L/2}(mT)/S_p \equiv S/S_p$ as a function of
$m$ at $\hbar \omega_D=Q_0$, $L=24$ and $Q_0/J=40$ for several initial Fock states. $S/S_p$ saturates to three distinct values showing signatures of weak fragmentation. (b) Plot of the half-chain entanglement $S_{L/2}/S_p$ for Floquet eigenstates as a function of their quasienergy $E_F$ for $\hbar \omega_D=Q_0=40J$ for $L=18$, showing two distinct clusters separated by a gap in $S$. For both plots, the initial states are chosen from the $W_{\ell}=1$ sector, and all energies are scaled in units of $J$.
See text for details.} \label{figen1}
\end{figure} 

Next, we compute the entanglement $S(mT)/S_p$ for the frequency $\hbar \omega_D=Q_0$ for which $H_F^{(1)}= H_a +2 H_B/\pi$. Here the driven system saturates to different long-time values of $S(mT)/S_p$. As shown in Fig.\ \ref{figen1}(a), for the majority of initial Fock states, $S(mT)/S_p \simeq 0.9$ for large $m$; however, for other initial Fock states, $S(mT)/S_p \to 0.55, 0.7$ in the large $m$ limit. This indicates the possibility of having a few dynamically disconnected fragments, one of which has a much larger HSD than the others. We note that the number of values to which $S(mT)$ saturates is much smaller than at $\omega_D=\omega_1^{\ast}$ shown in Fig.\ \ref{fig1}(a). This behavior is further supported by Fig.\ \ref{figen1}(b), where we plot $S/S_p$ of the Floquet eigenstates as a function of their quasienergy $E_F/J$. The plot shows two distinct clusters of eigenstates with different entanglement entropies which is consistent with the behavior shown in Fig.\ \ref{figen1}(a). We stress that we do not find an exponentially large number of dynamically disconnected sectors at this drive frequency unlike at $\omega_D=\omega_n^{\ast}$. 

\begin{figure}
\rotatebox{0}{\includegraphics*[width=0.99 \linewidth]{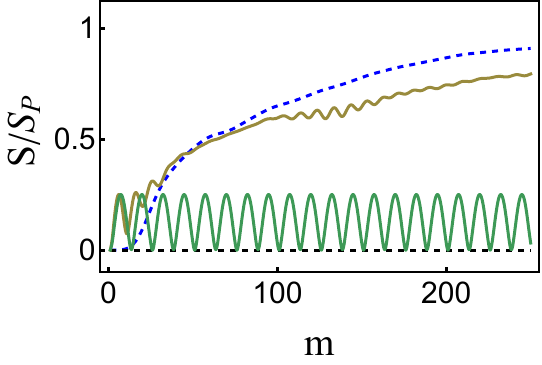}}
\caption{ Plot of $S_{L/2}(mT)/S_p \equiv S/S_p$ as a function of
$m$ at $\hbar \omega_D=Q_0$ (yellow solid line) $\hbar \omega_D=Q_0/2$ (green solid line) starting from the state $|\psi'_1\rangle$ which belongs to the largest fragment of $H'_a$. Similar plots starting from an initial frozen state $|\psi'_2\rangle$ is shown for $\hbar \omega_D=Q_0$ (blue dashed line) and $Q_0/2$ (black dashed line). For all
the plots, $L=24$, the initial states belong to the $W_{\ell}= \{\cdots 1,1,-1\cdots\}$ sector, $Q_0/J=40$, and all energies are scaled in units of $J$. See text for details. } \label{figen2}
\end{figure} 

Finally, we compute the entanglement entropy starting from initial states lying in the $W_{\ell}= \{\cdots 1,1,-1 \cdots\}$ sector. To this end, we begin by choosing an initial Fock state $|\psi'_1\rangle = | \cdots z y z z y 
z \cdots \rangle$ or $|\psi'_2\rangle =| \cdots x y x x 
y x\cdots \rangle$ and compute $|\psi'(mT)\rangle= U(mT,0) |\psi(0)\rangle$. The entanglement entropy $S(mT)$ is computed from the reduced density matrix corresponding to $|\psi'(mT)\rangle$ following the same procedure as described earlier for states in the $W_{\ell}=1$ sector.

The behavior of $S(mT)/S'_p$, where $S'_p= 0.45$ is the Page value of the sector, is shown in Fig.\ \ref{figen2}. For the initial state $|\psi'_1\rangle$ we find that $S/S_p$ increase towards the sector Page value if $\hbar \omega_D= Q_0$; a similar behavior (not shown in the figure) is found for the initial state $|\psi'_2\rangle$. We have numerically checked that such a growth of $S(mT)$ towards $S_p \simeq 0.45$ occurs for all $\hbar \omega_D \ne Q_0/(2n)$ (where $n \in Z$) for any randomly chosen initial Fock state. This 
supports the absence of fragmentation at $\omega_D=\omega'_n$ in this sector.

In contrast, for $\hbar \omega_D=Q_0/2$, these two states show qualitatively different behaviors of $S(mT)$. For an initial state $|\psi'_1\rangle$, $S(mT)$ oscillates between its initial value of zero
and the value of $\ln 2$. This can be understood by noting that the initial state $|\psi'_1\rangle$ belongs to the largest fragment of $H'_a$ (Eq.\ \ref{hamops2}). As shown in Sec.\ \ref{hsdw-1}, the states in the largest sector can be described as those of $L/3$ decoupled spins described by Pauli matrices $\tau_k$ living on the links of the chain. Constructing the half-chain essentially amounts to severing one such link which may lead to $S=0$ if the spin is in a $S_z$ eigenstate and $\ln 2$ if it is in a singlet state or a triplet state with $m_z=0$. These are therefore the two extremal possible value of $S$, and one finds an oscillation of $S(mT)$ between them. Such an oscillation is therefore a signature of the integrable nature of the largest fragment of $H'_a$. In contrast, the state $|\psi'_2\rangle$ is a frozen state, with $H'_a |\psi'_2\rangle=0$. Thus $S(mT)$ remains close to its initial zero value for a large prethermal time scale, as shown by the green dashed line of Fig.\ \ref{figen2}. Such a dependence of $S(mT)$ on the 
initial state is an indication of a prethermal emergent HSF at special drive frequencies and it shows a violation of ETH in finite-sized chains.

\subsection{Fidelity}
\label{fidel} 

\begin{figure}
\rotatebox{0}{\includegraphics*[width=0.49 \linewidth]{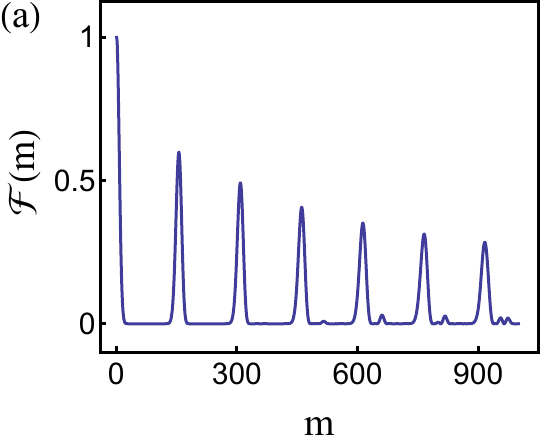}}
\rotatebox{0}{\includegraphics*[width=0.49 \linewidth]{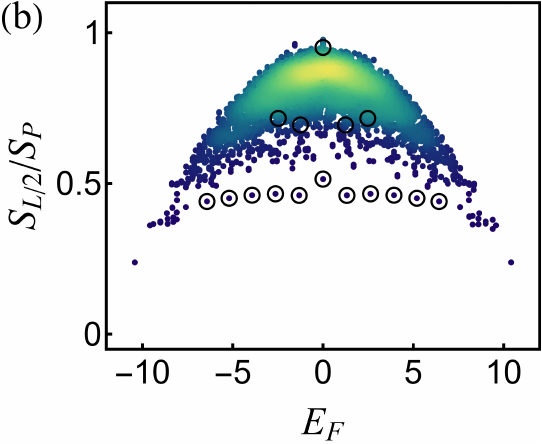}}
\rotatebox{0}{\includegraphics*[width=0.49 \linewidth]{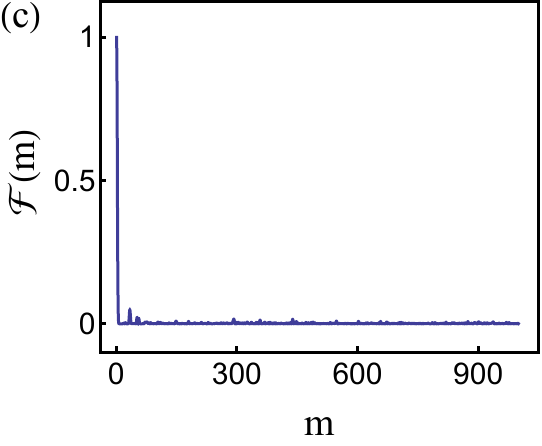}}
\rotatebox{0}{\includegraphics*[width=0.49 \linewidth]{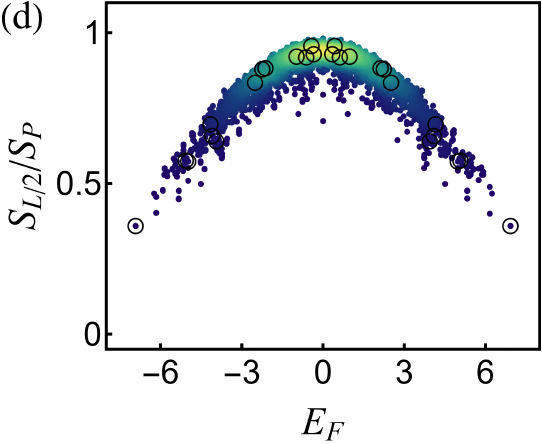}}
\caption{(a) Plot of ${\mathcal F}(mT)$ as a function of the number of drive cycles $m$ for $\hbar \omega_D/ Q_0=5$. ${\mathcal F}(mT)$ shows an oscillatory behavior; the amplitudes of these oscillations decrease with $m$. (b) Plot of $S_{L/2}/S_p$ for all Floquet eigenstates as a function of their quasienergy $E_F/J$ for $\hbar \omega_D/ Q_0=5$. The plot clearly shows the presence of quantum scar states with low $S_{L/2}$ separated from the usual thermal states with higher $S_{L/2}$; the circled states are the ones with high overlap with $|\psi(mT)\rangle$. (c) Same as in (a) but for $\hbar\omega_D/Q_0=1.25$ displaying ETH obeying rapid thermalization. (d) Same as in (b) but with $\hbar \omega_D/Q_0=1.25$. The plot indicates the absence of scars. For (a) and (c), $L=24$ while for (b) and(d) $L=18$. For all the plots, all states are chosen from the $W_{\ell}=1$ sector, $Q_0/J=40$, and all energies are scaled with $J$.} \label{fig4}
\end{figure} 

In this section, we compute the fidelity 
\begin{eqnarray} 
{\mathcal F}(mT) &=& |\langle \psi(mT)|\psi(0)\rangle| \label{fid1} 
\end{eqnarray}
starting from the initial state $|\psi(0)\rangle= |\psi_1\rangle$. We begin with the sector with all $W_{\ell}=1$. The dimension of this sector scales as $[(\sqrt{5}+1)/2]^L$ which indicates that the model has 
the same HSD as the PXP chain. Indeed, it has been shown that the $W_{\ell}=1$ sector of the model hosts quantum scars that have a large overlap with $|\psi_1\rangle$; the state $|\psi_1\rangle$ can be mapped to the Neel state of the PXP model using the transformation $|\cdots y_j x_{j+1 }\cdots \rangle \to |\cdots \uparrow_{\ell} \cdots \rangle$ and $|\cdots x_j y_{j+1} \cdots \rangle \to |\cdots \downarrow_{\ell} \cdots \rangle$, where $\ell$ denotes the link between the sites $j$ and $j+1$ \cite{sm2}. 

A plot of ${\mathcal F}(mT)$ as a function of $m$, shown in Fig.\ \ref{fig4}(a) for $\hbar \omega_D/Q_0=5$, shows an oscillatory behavior with distinct revivals; the amplitude of these oscillations decreases in time. This behavior constitutes a clear violation of ETH and can be attributed to quantum scars hosted by $H_F^{(1)}$. The presence of such scars can be understood by noting that for $\hbar \omega_d \gg Q_0$, $H_a$, $H_b$ and $H_c$ have almost the same amplitudes (Eq.\ \ref{flham1}); consequently, $H_F^{(1)} \simeq H_1$ which is known to host such scars \cite{sm2}. These scars, which have a large overlap with the state $|\psi(mT)\rangle$, can be clearly seen in Fig.\ \ref{fig4}(b). In contrast, at $\hbar \omega_D/Q_0=1.25$, ${\mathcal F}(mT)$ shows a quick thermalization consistent with ETH as seen in Fig.\ \ref{fig4}(c); concomitantly, a plot of $S/S_p$ as a function of the energy $E_F/J$ (Fig.\ \ref{fig4}(d)) shows that $H_F^{(1)}$ does not host scars for this drive frequency. This can be understood by noting that at this frequency, the amplitudes $H_a$, $H_{b}$ and $H_c$ are not identical; consequently, $H_F^{(1)}$ is not close to $H_1$ and it does not host quantum scars. Thus a reduction of the drive frequency at a fixed $Q_0$ (or equivalently increasing $Q_0/\omega_D$), leads to a crossover from a scar-dominated oscillatory behavior to ETH obeying rapid thermalization of the fidelity. 

\begin{figure}
\rotatebox{0}{\includegraphics*[width=0.49 \linewidth]{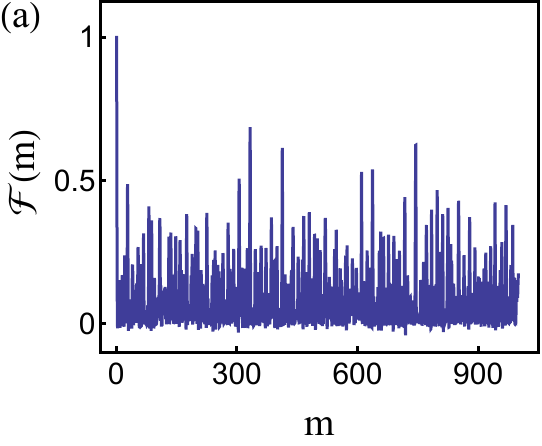}}
\rotatebox{0}{\includegraphics*[width=0.49 \linewidth]{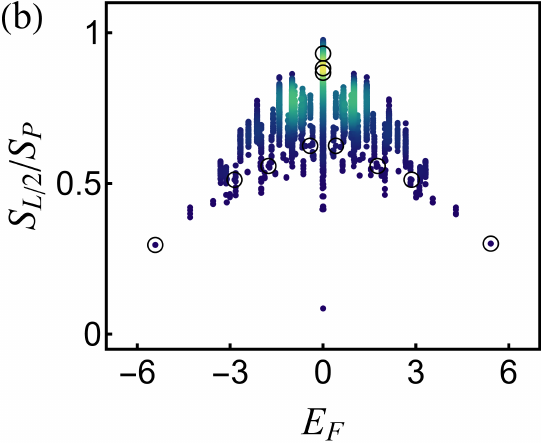}}
\caption{(a) Plot of ${\mathcal F}(mT)$ as a function of $m$ for $\omega_D=\omega_1^{\ast}= Q_0/(2\hbar)$ and $L=24$ showing strong finite-size fluctuations. (b) Plot of $S_{L/2}/S_p$ for all Floquet eigenstates for $L=18$ as a function of their quasienergy $E_F/J$ at $\omega_D=\omega_1^{\ast}$. For both plots, all states are chosen from the $W_{\ell}=1$ sector, $Q_0/J=40$, and all energies are scaled with $J$.} \label{fig5}
\end{figure} 

Upon lowering $\omega_D$ further, we reach $\omega_1^{\ast}$ for which $H_F^{(1)}= H_a$ and the corresponding Hilbert space is strongly fragmented. Here the fidelity depends on the initial state; our choice of $|\psi(0)\rangle=|\psi_1\rangle$ ensures that we address fidelity in the largest fragment. Since this fragment is ergodic, we find rapid thermalization. However, in this case, the HSD of the sector is much smaller compared to the total HSD leading to stronger finite-size effects as can be seen in Fig.\ \ref{fig5}(a). The fragmented structure of the states can be inferred from Fig.\ \ref{fig5}(b) which shows a plot of $S/S_p$ for all Floquet eigenstates as a function of their quasienergies $E_F/J$. The approximately fragmented structure of the Hilbert space can be inferred from the presence of a larger number of zero energy states; we have numerically verified that many of these have large overlaps with Fock states which indicates the presence of near frozen states in the spectrum. We note that the fragmentation here is approximate due to the presence of small but non-zero higher order terms in the Floquet Hamiltonian that do not support HSF. 

\begin{figure}
\rotatebox{0}{\includegraphics*[width=0.49 \linewidth]{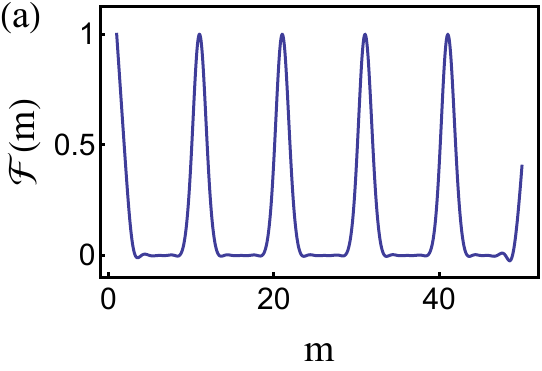}}
\rotatebox{0}{\includegraphics*[width=0.49 \linewidth]{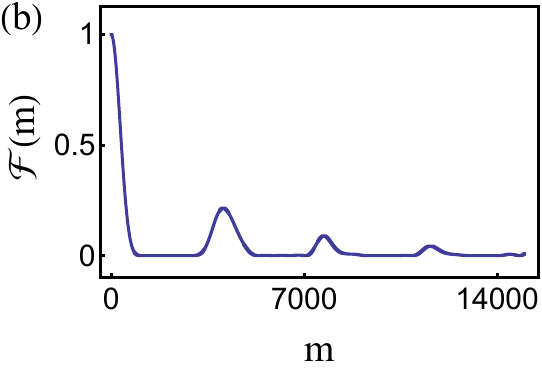}}
\caption{(a) Plot of ${\mathcal F}(mT)$ starting from the initial state $|\psi'_1\rangle$ as a function of the number of drive cycles $m$ for $\omega_D=\omega_1^{\ast}= Q_0/(2\hbar)$ shows an oscillatory behavior with repeated revivals. (b) Same plot but with initial state $|\psi'_2\rangle$ shows thermalization. For both plots, $L=24$, all initial stares are chosen from the $W_{\ell}=\{\cdots 1,1,-1\cdots\}$ sector, $Q_0/J=40$, and all energies are scaled with $J$.} \label{figfidw-1}
\end{figure} 

Finally, we consider the fidelity in the $W_{\ell}=\{\cdots 1,1,-1\cdots\}$ sector. Here for $\hbar \omega_D=Q_0/(2n)$, the behavior of ${\mathcal F}(mT)$ is qualitatively similar to the $W_{\ell}=1$ sector. For $\hbar \omega_D \gg Q_0$, $H_F^{'(1)} \to H_1$ and consequently ${\mathcal F}(mT)$ exhibits scar-induced oscillations similar to that in Fig.\ \ref{fig4}(a). As $\hbar \omega_D$ is lowered, we find rapid thermalization as seen in Fig.\ \ref{fig4}(c). 

However, the behavior of ${\mathcal F}(mT)$ at the special drive frequency $\omega_D=\omega_1^{\ast}$ is qualitatively different from its counterpart for the $W_{\ell}=1$ sector. This is shown in Fig.\ \ref{figfidw-1}. For the initial state $|\psi'_1\rangle$, the first-order Floquet Hamiltonian can be thought as $L/3$ decoupled systems given by Eq.\ \ref{oscham} each of which has two states with a quasienergy difference $2J$; consequently, ${\mathcal F}(mT)$ shows periodic revivals after $m^{\ast} = {\rm Int} (\pi/J T)$. For the parameters used for Fig.\ \ref{figfidw-1}(a), we find $\hbar \omega_D=20J$ which predicts $m^{\ast} \sim 10$; this matches the numerical result shown quite well. 

In contrast, for $|\psi(0)\rangle = |\psi'_2\rangle$ which is a frozen state of $H_F^{(1)}$, we find ${\mathcal F}(mT)$ displays slow dynamics as shown in Fig.\ \ref{figfidw-1}(b). The deviation of ${\mathcal F}(mT)$ from its initial value in this case occurs solely due to higher order terms in the Floquet Hamiltonian which is non-integrable. Thus ${\mathcal F}(mT)$ eventually reaches its ETH predicted thermal value $\simeq 0$; however, the thermalization is extremely slow since these higher order terms have small amplitudes for $Q_0/J\gg 1$. Such slow thermalization is a consequence of the slow dynamics of a frozen state due to higher-order terms in the Floquet Hamiltonian.

\subsection {Correlation functions}
\label{corr} 

\begin{figure}
\rotatebox{0}{\includegraphics*[width=0.49 \linewidth]{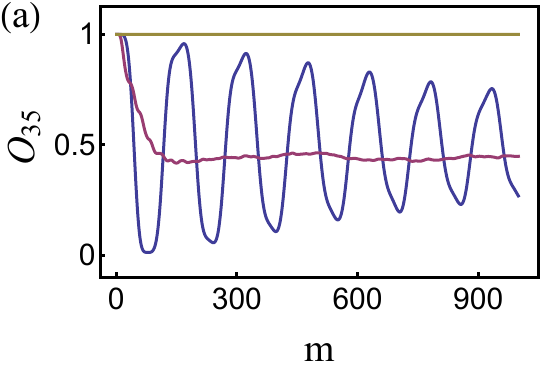}}
\rotatebox{0}{\includegraphics*[width=0.49 \linewidth]{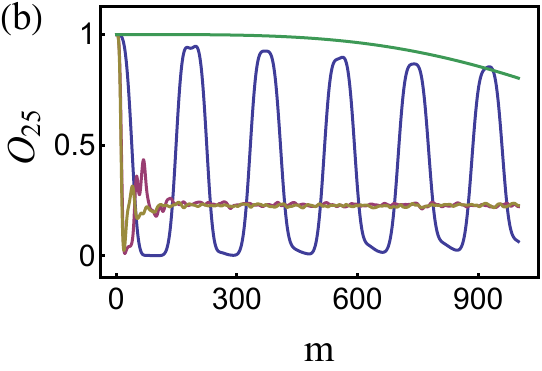}}
\caption{(a) Plot of $C_{24}$ starting from the initial state $|\psi_1\rangle$ in the $W_{\ell}=1$ sector as a function of the number of drive cycles $m$ for $\hbar \omega_D= 5Q_0$ (blue), $Q_0$ (magenta) and $Q_0/2$ showing scar-induced oscillations, ETH predicted thermalization, and near-constant behavior. (b) Plot of $C_{25}$ with initial state $|\psi'_1\rangle$ in the $W_{\ell}= \{\cdots 1,1,-1 \cdots\}$ sector and with the same color scheme as in (a). For both plots, $L=24$, $Q_0/J=40$, and all energies are scaled with $J$.} \label{figcorr}
\end{figure}

In this section, we study the equal-time correlation function 
\begin{eqnarray}
C_{j,j+n}(mT) \equiv C_0 &=& \langle \psi(mT)|\hat O_j \hat O_{j+n}|\psi(mT)\rangle, \nonumber\\
\hat O_j &=& (S_j^x S_{j+1}^y)^2, \label{correq1} 
\end{eqnarray}
as a function of $m$ for several drive frequencies. In what follows, we concentrate on $C_{24}$ for the $W_{\ell}=1$ sector and $C_{2 5}$ for the $W_{\ell}=\{\cdots 1,1,-1 \cdots\}$ sector; our numerical results shown in Fig.\ \ref{figcorr} shows qualitatively similar behaviors of these correlators in both sectors.

For both sectors at large drive frequencies $\hbar \omega_D \gg Q_0$, $H_F^{(1)} \sim H_0$; hence it supports quantum scars. For initial states, such as $|\psi_1\rangle$ in the $W_{\ell}=1$ sector and $|\psi'_1\rangle$ in the $W_{\ell}=\{\cdots 1,1,-1 \cdots\}$ sector, which have large overlaps with the scar states, $C_0$ shows oscillatory dynamics. This is shown in Figs.\ \ref{figcorr} (a) and (b) for $\hbar \omega_D= 5 Q_0$ (blue solid lines). Upon reducing the frequencies to $\hbar \omega_D= Q_0$, one sees a ETH predicted thermalization as shown in magenta lines in both panels. 
The intermediate region shows a gradual crossover between the oscillatory and the thermal behavior and a monotonic reduction in the thermalization time. 

At the special frequency $\hbar \omega_D= Q_0/2$, both the sectors show almost constant values of $C_0$ till a very large value of $m$. To understand this behavior we consider each sector separately. For the $W_{\ell}=1$ sector, for $\hbar \omega_D=Q_0/2$, we have $H_F^{(1)}= H_a$. It can be straightforwardly checked, using Eqs.\ \ref{hamops}, that $H_a$ connects a state with $|\cdots x_{j-1} y_{j} x_{j+1} y_{j+2} x_{j+3}\cdots \rangle$ to $|\cdots x_{j-1} z_{j} z_{j+1} y_{j+2} x_{j+3} \cdots \rangle$. Both these states are eigenstates of the operators $\hat O_j \hat O_{j+2}$. This indicates that $C_{j,j+2}$ for a fixed $j$ and $n=2$ will be pinned to its initial value. For the $W_{\ell}=\{\cdots 1,1,-1\cdots\}$ sector, one finds from Eq.\ \ref{hamops2} that $H'_a$ annihilates the initial state and the dynamics is completely frozen. Thus $C_{2 5}$ stays close to its original value. The slow deviation of $C_{25}$ from its initial value in the present case is a consequence of higher-order terms in the Floquet Hamiltonian.

\section{Discussion}
\label{diss} 

In this work, we have analyzed the dynamics of a periodically driven spin-one chain. Th model used is the well-known Kitaev model in one dimension. The key property of the model which makes this analysis possible is the presence of $Z_2$-valued conserved quantities, $W_{\ell}$, on every link $\ell$ of the chain. This fragments the Hilbert space of the spin-chain into sectors labeled by specific values of $W_{\ell}$
on each link $\ell$. The dimension of the largest (all $W_{\ell}=1$) and the second-largest ($W_{\ell}=\{\cdots 1,1,-1\cdots\}$) sectors are much smaller than the total HSD; this allows one to carry out ED on a finite-sized chain with $L\le 24$.

\begin{figure}
\rotatebox{0}{\includegraphics*[width=\linewidth]{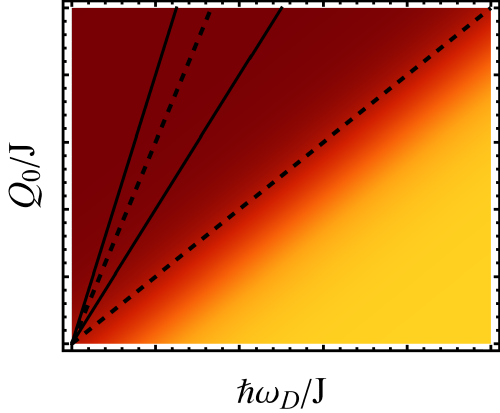}}
\caption{ Schematic phase diagram for the driven spin chain in the $Q_0-\omega_D$ plane. The yellow region shows scar-induced oscillations while the red region represents Floquet ETH predicted thermalization. Lowering $\omega_D$ leads to a gradual crossover between the two regions. The dotted lines correspond to $\hbar \omega_D= Q_0/(2n+1)$
(where $n=0,1,2,\cdots$); they are present only for the sector with all $W_{\ell}=1$ and represents weak HSF. They have no analogs in the $W_{\ell}=\{\cdots 1,1,-1\cdots\}$ sector. The solid lines at $\hbar \omega_D= Q_0/(2n)$ 
(where $n=1,2,3,\dots$) represent strong HSF for both sectors. See text for details.} \label{sch1}
\end{figure} 

We combine exact numerical results from such ED with analytical, perturbative, FPT-based calculation of the first-order Floquet Hamiltonian $H_F^{(1)}$ in the large drive amplitude regime to obtain our results. These are summarized in the schematic phase diagram shown in Fig.\ \ref{sch1}. For large $\hbar \omega_D/Q_0$, $H_F^{(1)}$ hosts quantum scars for both sectors. For the sector with all $W_{\ell}=1$, such scars are the same as those found for the PXP model; in both sectors, such scars lead to oscillatory dynamics of the fidelity and correlation functions. Upon reducing the drive frequency, such scar-induced dynamics is replaced by rapid, ETH predicted, thermalization in both sectors. This is shown schematically in Fig.\ \ref{sch1} as a crossover between yellow and red regions. For $\hbar \omega_D= Q_0/(2n+1)$, one finds a weak HSF for $W_{\ell}=1$. The number of Hilbert space fragments grows with
the system size but not exponentially. This is shown by dotted lines in Fig.\ \ref{sch1}. For the $W_{\ell}=\{\cdots 1,1,-1\cdots\}$ sector, no fragmentation occurs at these frequencies. 

For $\hbar \omega_D= Q_0/(2n)$, we find strong HSF for both sectors. For the sector with all $W_{\ell}=1$, the largest fragment is ergodic while for the $W_{\ell}=\{\cdots 1,1,-1\cdots\}$ sector, it is integrable. The presence of such fragmentation is most easily verified by a numerical computation of the half-chain entanglement $S(mT)$ for the $W_{\ell}=1$ sector; for large $m$, $S$ saturates to different values depending on the initial state. This feature shows violation of ETH and provides a clear indication of the fragmented structure of the Hilbert space. The integrable nature of the largest fragment in the $W_{\ell}= \{\cdots 1,1,-1\cdots\}$ sector shows up in an oscillatory behavior of $S$ (between the extremal values $S=0$ and $S=\ln 2$) 
for an initial state which belongs to this fragment. In addition, ${\mathcal F}$ shows perfect revivals for states in this fragment. These results provide concrete support to the 
analytic effective decoupled spin-half Hamiltonian (Eq.\ \ref{oscham}) obtained for this fragment. Furthermore, the correlation functions shows extremely slow dynamics at these 
frequencies for both sectors. Our numerical results, for both sectors, provide support to the analytic Floquet Hamiltonian obtained using FPT. 

To conclude, we have studied the Floquet dynamics of a spin-one chain using both FPT and ED. Our analysis based on this indicates the presence of a rich phase diagram for such dynamics supporting scar-induced oscillations, ETH predicted thermalization, and strong and weak HSF in different regimes of the drive frequency and amplitude. Our study opens up the possibility of studying random and quasiperiodic dynamics of the model which we leave as a subject of future study. 

\section{Acknowledgments} K.S. thanks
DST, India for support through the project JCB/2021/000030. D.S. thanks SERB, India for support through the project JBR/2920/000043.

\bibliography{refs}

@article{rev1,
  title = {Dynamics of a quantum phase transition and relaxation to a steady state},
  volume = {59},
  ISSN = {1460-6976},
  url = {http://dx.doi.org/10.1080/00018732.2010.514702},
  DOI = {10.1080/00018732.2010.514702},
  number = {6},
  journal = {Advances in Physics},
  publisher = {Informa UK Limited},
  author = {Dziarmaga,  Jacek},
  year = {2010},
  month = sep,
  pages = {1063–1189}
}

@article{rev2,
  title = {Colloquium: Nonequilibrium dynamics of closed interacting quantum systems},
  author = {Polkovnikov, Anatoli and Sengupta, Krishnendu and Silva, Alessandro and Vengalattore, Mukund},
  journal = {Rev. Mod. Phys.},
  volume = {83},
  issue = {3},
  pages = {863--883},
  numpages = {0},
  year = {2011},
  month = {Aug},
  publisher = {American Physical Society},
  doi = {10.1103/RevModPhys.83.863},
  url = {https://link.aps.org/doi/10.1103/RevModPhys.83.863}
}

@book{rev3,
  title = {Quantum Phase Transitions in Transverse Field Spin Models: From Statistical Physics to Quantum Information},
  ISBN = {9781107706057},
  url = {http://dx.doi.org/10.1017/CBO9781107706057},
  DOI = {10.1017/cbo9781107706057},
  publisher = {Cambridge University Press, Cambridge},
  author = {Dutta,  Amit and Aeppli,  Gabriel and Chakrabarti,  Bikas K. and Divakaran,  Uma and Rosenbaum,  Thomas F. and Sen,  Diptiman},
  year = {2015},
  month = jan 
}

@book{rev4,
  title = {Quantum Quenching,  Annealing and Computation},
  author = {Chandra, Anjan and Das, Arnab and Chakrabarti, Bikas},
  ISBN = {9783642114700},
  ISSN = {1616-6361},
  url = {http://dx.doi.org/10.1007/978-3-642-11470-0},
  DOI = {10.1007/978-3-642-11470-0},
  journal = {Lecture Notes in Physics},
  publisher = {Springer Berlin Heidelberg},
  year = {2010}
}

@article{rev5,
author = {Marin Bukov and Luca D'Alessio and Anatoli Polkovnikov},
title = {Universal high-frequency behavior of periodically driven systems: from dynamical stabilization to {F}loquet engineering},
journal = {Advances in Physics},
volume = {64},
number = {2},
pages = {139--226},
year = {2015},
publisher = {Taylor \& Francis},
doi = {10.1080/00018732.2015.1055918},
URL = {https://doi.org/10.1080/00018732.2015.1055918},
eprint = {https://doi.org/10.1080/00018732.2015.1055918}

}

@article{rev6,
title = {Many-body energy localization transition in periodically driven systems},
journal = {Annals of Physics},
volume = {333},
pages = {19-33},
year = {2013},
issn = {0003-4916},
doi = {https://doi.org/10.1016/j.aop.2013.02.011},
url = {https://www.sciencedirect.com/science/article/pii/S0003491613000389},
author = {Luca D’Alessio and Anatoli Polkovnikov}
}

@article{rev7,
author = {Luca D'Alessio and Yariv Kafri and Anatoli Polkovnikov and Marcos Rigol},
title = {From quantum chaos and eigenstate thermalization to statistical mechanics and thermodynamics},
journal = {Advances in Physics},
volume = {65},
number = {3},
pages = {239--362},
year = {2016},
publisher = {Taylor \& Francis},
doi = {10.1080/00018732.2016.1198134},
URL = {https://doi.org/10.1080/00018732.2016.1198134},
eprint = {https://doi.org/10.1080/00018732.2016.1198134}

}

@article{rev8,
  title = {Landau–{Z}ener–{S}t\"{u}ckelberg interferometry},
  volume = {492},
  ISSN = {0370-1573},
  url = {http://dx.doi.org/10.1016/j.physrep.2010.03.002},
  DOI = {10.1016/j.physrep.2010.03.002},
  number = {1},
  journal = {Physics Reports},
  publisher = {Elsevier BV},
  author = {Shevchenko,  S.N. and Ashhab,  S. and Nori,  Franco},
  year = {2010},
  month = jul,
  pages = {1-30}
}

@article{rev9,
  title = {Floquet Engineering of Quantum Materials},
  volume = {10},
  ISSN = {1947-5462},
  url = {http://dx.doi.org/10.1146/annurev-conmatphys-031218-013423},
  DOI = {10.1146/annurev-conmatphys-031218-013423},
  number = {1},
  journal = {Annual Review of Condensed Matter Physics},
  publisher = {Annual Reviews},
  author = {Oka,  Takashi and Kitamura,  Sota},
  year = {2019},
  month = mar,
  pages = {387–408}
}

@article{rev10,
  title = {The {M}agnus expansion and some of its applications},
  volume = {470},
  ISSN = {0370-1573},
  url = {http://dx.doi.org/10.1016/j.physrep.2008.11.001},
  DOI = {10.1016/j.physrep.2008.11.001},
  number = {5–6},
  journal = {Physics Reports},
  publisher = {Elsevier BV},
  author = {Blanes,  S. and Casas,  F. and Oteo,  J.A. and Ros,  J.},
  year = {2009},
  month = jan,
  pages = {151–238}
}

@article{rev11,
  title = {Colloquium: Atomic quantum gases in periodically driven optical lattices},
  author = {Eckardt, Andr\'e},
  journal = {Rev. Mod. Phys.},
  volume = {89},
  issue = {1},
  pages = {011004},
  numpages = {30},
  year = {2017},
  month = {Mar},
  publisher = {American Physical Society},
  doi = {10.1103/RevModPhys.89.011004},
  url = {https://link.aps.org/doi/10.1103/RevModPhys.89.011004}
}

@article{rev12,
doi = {10.1088/1361-648X/ac1b61},
url = {https://dx.doi.org/10.1088/1361-648X/ac1b61},
year = {2021},
month = {aug},
publisher = {IOP Publishing},
volume = {33},
number = {44},
pages = {443003},
author = {Sen, Arnab and Sen, Diptiman and Sengupta, K},
title = {Analytic approaches to periodically driven closed quantum systems: methods and applications},
journal = {Journal of Physics: Condensed Matter}
}

@article{rev13,
  title = {Many-body physics with ultracold gases},
  author = {Bloch, Immanuel and Dalibard, Jean and Zwerger, Wilhelm},
  journal = {Rev. Mod. Phys.},
  volume = {80},
  issue = {3},
  pages = {885--964},
  numpages = {0},
  year = {2008},
  month = {Jul},
  publisher = {American Physical Society},
  doi = {10.1103/RevModPhys.80.885},
  url = {https://link.aps.org/doi/10.1103/RevModPhys.80.885}
}

@article{rev14,
  title = {Quantum simulation of the {H}ubbard model with ultracold fermions in optical lattices},
  volume = {19},
  ISSN = {1878-1535},
  url = {http://dx.doi.org/10.1016/j.crhy.2018.10.013},
  DOI = {10.1016/j.crhy.2018.10.013},
  number = {6},
  journal = {Comptes Rendus. Physique},
  publisher = {Cellule MathDoc/Centre Mersenne},
  author = {Tarruell,  Leticia and Sanchez-Palencia,  Laurent},
  year = {2018},
  month = sep,
  pages = {365–393}
}

@article{rev15,
  title = {Quantum and classical Floquet prethermalization},
  volume = {454},
  ISSN = {0003-4916},
  url = {http://dx.doi.org/10.1016/j.aop.2023.169297},
  DOI = {10.1016/j.aop.2023.169297},
  journal = {Annals of Physics},
  publisher = {Elsevier BV},
  author = {Ho,  Wen Wei and Mori,  Takashi and Abanin,  Dmitry A. and Dalla Torre,  Emanuele G.},
  year = {2023},
  month = jul,
  pages = {169297}
}

@article{rev16,
  title = {Thermalization and prethermalization in isolated quantum systems: a theoretical overview},
  volume = {51},
  ISSN = {1361-6455},
  url = {http://dx.doi.org/10.1088/1361-6455/aabcdf},
  DOI = {10.1088/1361-6455/aabcdf},
  number = {11},
  journal = {Journal of Physics B},
  publisher = {IOP Publishing},
  author = {Mori,  Takashi and Ikeda,  Tatsuhiko N and Kaminishi,  Eriko and Ueda,  Masahito},
  year = {2018},
  month = may,
  pages = {112001}
}

@article{rev17,
  title = {Emergent symmetries in prethermal phases of periodically driven quantum systems},
  volume = {37},
  ISSN = {1361-648X},
  url = {http://dx.doi.org/10.1088/1361-648X/ada860},
  DOI = {10.1088/1361-648x/ada860},
  number = {13},
  journal = {Journal of Physics: Condensed Matter},
  publisher = {IOP Publishing},
  author = {Banerjee,  Tista and Sengupta,  K},
  year = {2025},
  month = feb,
  pages = {133002}
}

@article{exp1,
  title = {A quantum gas microscope for detecting single atoms in a {H}ubbard-regime optical lattice},
  volume = {462},
  ISSN = {1476-4687},
  url = {http://dx.doi.org/10.1038/nature08482},
  DOI = {10.1038/nature08482},
  number = {7269},
  journal = {Nature},
  publisher = {Springer Science and Business Media LLC},
  author = {Bakr,  Waseem S. and Gillen,  Jonathon I. and Peng,  Amy and F\"{o}lling,  Simon and Greiner,  Markus},
  year = {2009},
  month = nov,
  pages = {74}
}

@article{exp2,
  title = {Probing the Superfluid–to–{M}ott Insulator Transition at the Single-Atom Level},
  volume = {329},
  ISSN = {1095-9203},
  url = {http://dx.doi.org/10.1126/science.1192368},
  DOI = {10.1126/science.1192368},
  number = {5991},
  journal = {Science},
  publisher = {American Association for the Advancement of Science (AAAS)},
  author = {Bakr,  W. S. and Peng,  A. and Tai,  M. E. and Ma,  R. and Simon,  J. and Gillen,  J. I. and F\"{o}lling,  S. and Pollet,  L. and Greiner,  M.},
  year = {2010},
  month = jul,
  pages = {547}
}

@article{exp3,
  title = {Probing many-body dynamics on a 51-atom quantum simulator},
  volume = {551},
  ISSN = {1476-4687},
  url = {http://dx.doi.org/10.1038/nature24622},
  DOI = {10.1038/nature24622},
  number = {7682},
  journal = {Nature},
  publisher = {Springer Science and Business Media LLC},
  author = {Bernien,  Hannes and Schwartz,  Sylvain and Keesling,  Alexander and Levine,  Harry and Omran,  Ahmed and Pichler,  Hannes and Choi,  Soonwon and Zibrov,  Alexander S. and Endres,  Manuel and Greiner,  Markus and Vuletić,  Vladan and Lukin,  Mikhail D.},
  year = {2017},
  month = nov,
  pages = {579}
}

@article{exp4,
  title = {High-Fidelity Control and Entanglement of {R}ydberg-Atom Qubits},
  author = {Levine, Harry and Keesling, Alexander and Omran, Ahmed and Bernien, Hannes and Schwartz, Sylvain and Zibrov, Alexander S. and Endres, Manuel and Greiner, Markus and Vuleti\ifmmode \acute{c}\else \'{c}\fi{}, Vladan and Lukin, Mikhail D.},
  journal = {Phys. Rev. Lett.},
  volume = {121},
  issue = {12},
  pages = {123603},
  numpages = {6},
  year = {2018},
  month = {Sep},
  publisher = {American Physical Society},
  doi = {10.1103/PhysRevLett.121.123603},
  url = {https://link.aps.org/doi/10.1103/PhysRevLett.121.123603}
}

@article{exp5,
  title = {Controlling quantum many-body dynamics in driven {R}ydberg atom arrays},
  author = {D. Bluvstein  and A. Omran  and H. Levine  and A. Keesling  and G. Semeghini  and S. Ebadi  and T. T. Wang  and A. A. Michailidis  and N. Maskara  and W. W. Ho  and S. Choi  and M. Serbyn  and M. Greiner  and V. Vuletić  and M. D. Lukin },
  journal = {Science},
  volume = {371},
  number = {6536},
  pages = {1355-1359},
  year = {2021},
  doi = {10.1126/science.abg2530},
  URL = {https://www.science.org/doi/abs/10.1126/science.abg2530}
}

@article{exp6,
  title={Quantum coarsening and collective dynamics on a programmable simulator},
  author={Manovitz, Tom and Li, Sophie H and Ebadi, Sepehr and Samajdar, Rhine and Geim, Alexandra A and Evered, Simon J and Bluvstein, Dolev and Zhou, Hengyun and Koyluoglu, Nazli Ugur and Feldmeier, Johannes and Dolgirev, Pavel E and Maskara, Nishad and Kalinowski, Marcin and  Sachdev, Subir and Huse, David A and Greiner, Markus and Vuletić, Vladan and Lukin, Mikhail D},
  journal={Nature},
  volume={638},
  number={8049},
  pages={86--92},
  year={2025},
  publisher={Nature Publishing Group UK London},
   doi = {10.1103/w1cp-l5vq},
  url = {https://www.nature.com/articles/s41586-024-08353-5}
}

@article{exp7,
  title = {Observation of Anomalous Information Scrambling in a {R}ydberg Atom Array},
  author = {Liang, Xinhui and Yue, Zongpei and Chao, Yu-Xin and Hua, Zhen-Xing and Lin, Yige and Tey, Meng Khoon and You, Li},
  journal = {Phys. Rev. Lett.},
  volume = {135},
  issue = {5},
  pages = {050201},
  numpages = {7},
  year = {2025},
  month = {Jul},
  publisher = {American Physical Society},
  doi = {10.1103/w1cp-l5vq},
}

@article{dynloc1,
  title = {From dynamical localization to bunching in interacting Floquet systems},
  volume = {5},
  ISSN = {2542-4653},
  url = {http://dx.doi.org/10.21468/SciPostPhys.5.2.017},
  DOI = {10.21468/scipostphys.5.2.017},
  number = {2},
  journal = {SciPost Physics},
  publisher = {Stichting SciPost},
  author = {Baum,  Yuval and van Nieuwenburg,  Everard and Refael,  Gil},
  year = {2018},
  month = aug 
}

@article{dynloc2,
  title = {Absence of dynamical localization in interacting driven systems},
  volume = {3},
  ISSN = {2542-4653},
  url = {http://dx.doi.org/10.21468/SciPostPhys.3.4.029},
  DOI = {10.21468/scipostphys.3.4.029},
  number = {4},
  journal = {SciPost Physics},
  publisher = {Stichting SciPost},
  author = {Luitz,  David J. and Bar Lev,  Yevgeny and Lazarides,  Achilleas},
  year = {2017},
  month = oct 
}

@article{dynloc3,
  title = {Effects of interactions on periodically driven dynamically localized systems},
  volume = {95},
  ISSN = {2469-9969},
  url = {http://dx.doi.org/10.1103/PhysRevB.95.014305},
  DOI = {10.1103/physrevb.95.014305},
  number = {1},
  pages={014305},
  journal = {Phys. Rev. B},
  publisher = {American Physical Society (APS)},
  author = {Agarwala,  Adhip and Sen,  Diptiman},
  year = {2017},
  month = jan 
}

@Article{dynloc4,
	title={{Dynamical localization and slow thermalization in a class of disorder-free periodically driven one-dimensional interacting systems}},
	author={Sreemayee Aditya and Diptiman Sen},
	journal={SciPost Phys. Core},
	volume={6},
	pages={083},
	year={2023},
	publisher={SciPost},
	doi={10.21468/SciPostPhysCore.6.4.083},
	url={https://scipost.org/10.21468/SciPostPhysCore.6.4.083},
}

@article{dynloc5,
  title = {Dynamical localization in a chain of hard core bosons under periodic driving},
  author = {Nag, Tanay and Roy, Sthitadhi and Dutta, Amit and Sen, Diptiman},
  journal = {Phys. Rev. B},
  volume = {89},
  issue = {16},
  pages = {165425},
  numpages = {5},
  year = {2014},
  month = {Apr},
  publisher = {American Physical Society},
  doi = {10.1103/PhysRevB.89.165425},
  url = {https://link.aps.org/doi/10.1103/PhysRevB.89.165425}
}

@article{dynloc6,
  title = {Floquet engineering of low-energy dispersions and dynamical localization in a periodically kicked three-band system},
  volume = {104},
  ISSN = {2469-9969},
  url = {http://dx.doi.org/10.1103/PhysRevB.104.174308},
  DOI = {10.1103/physrevb.104.174308},
  number = {17},
  pages={174308},
  journal = {Phys. Rev. B},
  publisher = {American Physical Society (APS)},
  author = {Tamang,  Lakpa and Nag,  Tanay and Biswas,  Tutul},
  year = {2021},
  month = nov 
}

@article{dynloc7,
  title = {Many-body dynamical localization in the kicked {B}ose-{H}ubbard chain},
  volume = {101},
  ISSN = {2469-9969},
  url = {http://dx.doi.org/10.1103/PhysRevB.101.064302},
  DOI = {10.1103/physrevb.101.064302},
  pages={064302},
  number = {6},
  journal = {Phys. Rev. B},
  publisher = {American Physical Society (APS)},
  author = {Fava,  Michele and Fazio,  Rosario and Russomanno,  Angelo},
  year = {2020},
  month = feb 
}

@article{dyntran1,
  title = {Dynamical Quantum Phase Transitions in the Transverse-Field Ising Model},
  author = {Heyl, M. and Polkovnikov, A. and Kehrein, S.},
  journal = {Phys. Rev. Lett.},
  volume = {110},
  issue = {13},
  pages = {135704},
  numpages = {5},
  year = {2013},
  month = {Mar},
  publisher = {American Physical Society},
  doi = {10.1103/PhysRevLett.110.135704},
  url = {https://link.aps.org/doi/10.1103/PhysRevLett.110.135704}
}

@article{dyntran2,
  title = {Observing Dynamical Quantum Phase Transitions through Quasilocal String Operators},
  author = {Bandyopadhyay, Souvik and Polkovnikov, Anatoli and Dutta, Amit},
  journal = {Phys. Rev. Lett.},
  volume = {126},
  issue = {20},
  pages = {200602},
  numpages = {6},
  year = {2021},
  month = {May},
  publisher = {American Physical Society},
  doi = {10.1103/PhysRevLett.126.200602},
  url = {https://link.aps.org/doi/10.1103/PhysRevLett.126.200602}
}

@article{dyntran3,
  title = {Dynamical quantum phase transitions in extended transverse Ising models},
  author = {Bhattacharjee, Sourav and Dutta, Amit},
  journal = {Phys. Rev. B},
  volume = {97},
  issue = {13},
  pages = {134306},
  numpages = {8},
  year = {2018},
  month = {Apr},
  publisher = {American Physical Society},
  doi = {10.1103/PhysRevB.97.134306},
  url = {https://link.aps.org/doi/10.1103/PhysRevB.97.134306}
}

@article{dyntran4,
doi = {10.1088/1361-6633/aaaf9a},
url = {https://doi.org/10.1088/1361-6633/aaaf9a},
year = {2018},
month = {apr},
publisher = {IOP Publishing},
volume = {81},
number = {5},
pages = {054001},
author = {Heyl, Markus},
title = {Dynamical quantum phase transitions: a review},
journal = {Reports on Progress in Physics}
}

@article{dyntran5,
  title = {Entanglement generation in periodically driven integrable systems: Dynamical phase transitions and steady state},
  author = {Sen, Arnab and Nandy, Sourav and Sengupta, K.},
  journal = {Phys. Rev. B},
  volume = {94},
  issue = {21},
  pages = {214301},
  numpages = {16},
  year = {2016},
  month = {Dec},
  publisher = {American Physical Society},
  doi = {10.1103/PhysRevB.94.214301},
  url = {https://link.aps.org/doi/10.1103/PhysRevB.94.214301}
}

@article{dyntran6,
  title = {Dynamical relaxation of correlators in periodically driven integrable quantum systems},
  author = {Aditya, Sreemayee and Samanta, Sutapa and Sen, Arnab and Sengupta, K. and Sen, Diptiman},
  journal = {Phys. Rev. B},
  volume = {105},
  issue = {10},
  pages = {104303},
  numpages = {14},
  year = {2022},
  month = {Mar},
  publisher = {American Physical Society},
  doi = {10.1103/PhysRevB.105.104303},
  url = {https://link.aps.org/doi/10.1103/PhysRevB.105.104303}
}

@article{dyntran7,
  title = {Periodically driven integrable systems with long-range pair potentials},
  volume = {51},
  ISSN = {1751-8121},
  url = {http://dx.doi.org/10.1088/1751-8121/aaced6},
  DOI = {10.1088/1751-8121/aaced6},
  number = {33},
  journal = {Journal of Physics A},
  publisher = {IOP Publishing},
  author = {Nandy,  Sourav and Sengupta,  K and Sen,  Arnab},
  year = {2018},
  month = jul,
  pages = {334002}

}

@article{dyntran8,
  title = {Dynamical crossover behavior in the relaxation of quenched quantum many-body systems},
  author = {Makki, Aamir Ahmad and Bandyopadhyay, Souvik and Maity, Somnath and Dutta, Amit},
  journal = {Phys. Rev. B},
  volume = {105},
  issue = {5},
  pages = {054301},
  numpages = {12},
  year = {2022},
  month = {Feb},
  publisher = {American Physical Society},
  doi = {10.1103/PhysRevB.105.054301},
  url = {https://link.aps.org/doi/10.1103/PhysRevB.105.054301}
}

@article{dynfr1,
  title = {Exotic freezing of response in a quantum many-body system},
  author = {Das, Arnab},
  journal = {Phys. Rev. B},
  volume = {82},
  issue = {17},
  pages = {172402},
  numpages = {4},
  year = {2010},
  month = {Nov},
  publisher = {American Physical Society},
  doi = {10.1103/PhysRevB.82.172402},
  url = {https://link.aps.org/doi/10.1103/PhysRevB.82.172402}
}

@article{dynfr2,
  title = {Freezing a quantum magnet by repeated quantum interference: An experimental realization},
  author = {Hegde, Swathi S. and Katiyar, Hemant and Mahesh, T. S. and Das, Arnab},
  journal = {Phys. Rev. B},
  volume = {90},
  issue = {17},
  pages = {174407},
  numpages = {7},
  year = {2014},
  month = {Nov},
  publisher = {American Physical Society},
  doi = {10.1103/PhysRevB.90.174407},
  url = {https://link.aps.org/doi/10.1103/PhysRevB.90.174407}
}

@article{dynfr3,
  title = {Dynamics-induced freezing of strongly correlated ultracold bosons},
  volume = {100},
  ISSN = {1286-4854},
  url = {http://dx.doi.org/10.1209/0295-5075/100/60007},
  DOI = {10.1209/0295-5075/100/60007},
  number = {6},
  journal = {EPL (Europhysics Letters)},
  publisher = {IOP Publishing},
  author = {Mondal,  S. and Pekker,  D. and Sengupta,  K.},
  year = {2012},
  month = dec,
  pages = {60007}
}

@article{dynfr4,
  title = {Dynamical Freezing in Exactly Solvable Models of Driven Chaotic Quantum Dots},
  volume = {134},
  ISSN = {1079-7114},
  url = {http://dx.doi.org/10.1103/ggk3-6cf8},
  DOI = {10.1103/ggk3-6cf8},
  number = {22},
  journal = {Phys. Rev. Lett.},
  publisher = {American Physical Society (APS)},
  author = {Guo,  Haoyu and Mukherjee,  Rohit and Chowdhury,  Debanjan},
  year = {2025},
  month = jun 
}

@article{dynfr5,
  title = {Dynamic freezing and defect suppression in the tilted one-dimensional {B}ose-{H}ubbard model},
  volume = {90},
  ISSN = {1550-235X},
  url = {http://dx.doi.org/10.1103/PhysRevB.90.184303},
  pages={184303},
  DOI = {10.1103/physrevb.90.184303},
  number = {18},
  journal = {Phys. Rev. B},
  publisher = {American Physical Society (APS)},
  author = {Divakaran,  U. and Sengupta,  K.},
  year = {2014},
  month = nov 
}

@article{dynfr6,
  title = {Dynamical Freezing and Scar Points in Strongly Driven {F}loquet Matter: Resonance vs Emergent Conservation Laws},
  author = {Haldar, Asmi and Sen, Diptiman and Moessner, Roderich and Das, Arnab},
  journal = {Phys. Rev. X},
  volume = {11},
  issue = {2},
  pages = {021008},
  numpages = {25},
  year = {2021},
  month = {Apr},
  publisher = {American Physical Society},
  doi = {10.1103/PhysRevX.11.021008},
  url = {https://link.aps.org/doi/10.1103/PhysRevX.11.021008}
}

@article{dynfr7,
  title = {Counterdiabatic Route to Entanglement Steering and Dynamical Freezing in the {F}loquet Lipkin-Meshkov-Glick Model},
  author = {Gangopadhay, Nakshatra and Choudhury, Sayan},
  journal = {Phys. Rev. Lett.},
  volume = {135},
  issue = {2},
  pages = {020407},
  numpages = {7},
  year = {2025},
  month = {Jul},
  publisher = {American Physical Society},
  doi = {10.1103/bzcf-gm89},
  url = {https://link.aps.org/doi/10.1103/bzcf-gm89}
}

@article{tcrev1,
  title = {Time crystals in periodically driven systems},
  volume = {71},
  ISSN = {1945-0699},
  url = {http://dx.doi.org/10.1063/PT.3.4020},
  DOI = {10.1063/pt.3.4020},
  number = {9},
  journal = {Physics Today},
  publisher = {AIP Publishing},
  author = {Yao,  Norman Y. and Nayak,  Chetan},
  year = {2018},
  month = sep,
  pages = {40–47}
}

@misc{tcrev2,
      title={A Brief History of Time Crystals}, 
      author={Vedika Khemani and Roderich Moessner and S. L. Sondhi},
      year={2019},
      eprint={1910.10745},
      archivePrefix={arXiv},
      primaryClass={cond-mat.str-el},
      url={https://arxiv.org/abs/1910.10745}, 
}

@article{tcrev3,
  title = {Discrete Time Crystals},
  volume = {11},
  ISSN = {1947-5462},
  url = {http://dx.doi.org/10.1146/annurev-conmatphys-031119-050658},
  DOI = {10.1146/annurev-conmatphys-031119-050658},
  number = {1},
  journal = {Annual Review of Condensed Matter Physics},
  publisher = {Annual Reviews},
  author = {Else,  Dominic V. and Monroe,  Christopher and Nayak,  Chetan and Yao,  Norman Y.},
  year = {2020},
  month = mar,
  pages = {467–499}
}

@article{tcrev4,
  title = {Colloquium
: Quantum and classical discrete time crystals},
  volume = {95},
  ISSN = {1539-0756},
  url = {http://dx.doi.org/10.1103/RevModPhys.95.031001},
  DOI = {10.1103/revmodphys.95.031001},
  number = {3},
  pages={031001},
  journal = {Rev. Mod. Phys.},
  publisher = {American Physical Society (APS)},
  author = {Zaletel,  Michael P. and Lukin,  Mikhail and Monroe,  Christopher and Nayak,  Chetan and Wilczek,  Frank and Yao,  Norman Y.},
  year = {2023},
  month = jul 
}

@article{tcrev5,
  title = {Time crystals: a review},
  volume = {81},
  ISSN = {1361-6633},
  url = {http://dx.doi.org/10.1088/1361-6633/aa8b38},
  DOI = {10.1088/1361-6633/aa8b38},
  number = {1},
  journal = {Reports on Progress in Physics},
  publisher = {IOP Publishing},
  author = {Sacha,  Krzysztof and Zakrzewski,  Jakub},
  year = {2017},
  month = nov,
  pages = {016401}
}

@article{tc1,
  title = {Absolute stability and spatiotemporal long-range order in {F}loquet systems},
  author = {von Keyserlingk, C. W. and Khemani, Vedika and Sondhi, S. L.},
  journal = {Phys. Rev. B},
  volume = {94},
  issue = {8},
  pages = {085112},
  numpages = {11},
  year = {2016},
  month = {Aug},
  publisher = {American Physical Society},
  doi = {10.1103/PhysRevB.94.085112},
  url = {https://link.aps.org/doi/10.1103/PhysRevB.94.085112}
}

@article{tc2,
  title = {Equilibration and order in quantum {F}loquet matter},
  volume = {13},
  ISSN = {1745-2481},
  url = {http://dx.doi.org/10.1038/nphys4106},
  DOI = {10.1038/nphys4106},
  number = {5},
  journal = {Nature Physics},
  publisher = {Springer Science and Business Media LLC},
  author = {Moessner,  R. and Sondhi,  S. L.},
  year = {2017},
  month = apr,
  pages = {424}
}

@article{tc3,
  title = {{F}loquet Time Crystals},
  author = {Else, Dominic V. and Bauer, Bela and Nayak, Chetan},
  journal = {Phys. Rev. Lett.},
  volume = {117},
  issue = {9},
  pages = {090402},
  numpages = {5},
  year = {2016},
  month = {Aug},
  publisher = {American Physical Society},
  doi = {10.1103/PhysRevLett.117.090402},
  url = {https://link.aps.org/doi/10.1103/PhysRevLett.117.090402}
}

@article{tc4,
  title = {Prethermal Phases of Matter Protected by Time-Translation Symmetry},
  author = {Else, Dominic V. and Bauer, Bela and Nayak, Chetan},
  journal = {Phys. Rev. X},
  volume = {7},
  issue = {1},
  pages = {011026},
  numpages = {21},
  year = {2017},
  month = {Mar},
  publisher = {American Physical Society},
  doi = {10.1103/PhysRevX.7.011026},
  url = {https://link.aps.org/doi/10.1103/PhysRevX.7.011026}
}

@article{tc5,
  title = {Discrete Time Crystals: Rigidity, Criticality, and Realizations},
  author = {Yao, N. Y. and Potter, A. C. and Potirniche, I.-D. and Vishwanath, A.},
  journal = {Phys. Rev. Lett.},
  volume = {118},
  issue = {3},
  pages = {030401},
  numpages = {6},
  year = {2017},
  month = {Jan},
  publisher = {American Physical Society},
  doi = {10.1103/PhysRevLett.118.030401},
  url = {https://link.aps.org/doi/10.1103/PhysRevLett.118.030401}
}

@article{topo1,
  title = {Topological characterization of periodically driven quantum systems},
  author = {Kitagawa, Takuya and Berg, Erez and Rudner, Mark and Demler, Eugene},
  journal = {Phys. Rev. B},
  volume = {82},
  issue = {23},
  pages = {235114},
  numpages = {12},
  year = {2010},
  month = {Dec},
  publisher = {American Physical Society},
  doi = {10.1103/PhysRevB.82.235114},
  url = {https://link.aps.org/doi/10.1103/PhysRevB.82.235114}
}

@article{topo2,
  title = {{F}loquet topological insulator in semiconductor quantum wells},
  volume = {7},
  ISSN = {1745-2481},
  url = {http://dx.doi.org/10.1038/nphys1926},
  DOI = {10.1038/nphys1926},
  number = {6},
  journal = {Nature Physics},
  publisher = {Springer Science and Business Media LLC},
  author = {Lindner,  Netanel H. and Refael,  Gil and Galitski,  Victor},
  year = {2011},
  month = mar,
  pages = {490}
}

@article{topo3,
  title = {Transport properties of nonequilibrium systems under the application of light: Photoinduced quantum Hall insulators without Landau levels},
  author = {Kitagawa, Takuya and Oka, Takashi and Brataas, Arne and Fu, Liang and Demler, Eugene},
  journal = {Phys. Rev. B},
  volume = {84},
  issue = {23},
  pages = {235108},
  numpages = {13},
  year = {2011},
  month = {Dec},
  publisher = {American Physical Society},
  doi = {10.1103/PhysRevB.84.235108},
  url = {https://link.aps.org/doi/10.1103/PhysRevB.84.235108}
}

@article{topo4,
  title = {{F}loquet generation of Majorana end modes and topological invariants},
  author = {Thakurathi, Manisha and Patel, Aavishkar A. and Sen, Diptiman and Dutta, Amit},
  journal = {Phys. Rev. B},
  volume = {88},
  issue = {15},
  pages = {155133},
  numpages = {13},
  year = {2013},
  month = {Oct},
  publisher = {American Physical Society},
  doi = {10.1103/PhysRevB.88.155133},
  url = {https://link.aps.org/doi/10.1103/PhysRevB.88.155133}
}

@article{topo5,
  title = {Effective Theory of {F}loquet Topological Transitions},
  author = {Kundu, Arijit and Fertig, H. A. and Seradjeh, Babak},
  journal = {Phys. Rev. Lett.},
  volume = {113},
  issue = {23},
  pages = {236803},
  numpages = {5},
  year = {2014},
  month = {Dec},
  publisher = {American Physical Society},
  doi = {10.1103/PhysRevLett.113.236803},
  url = {https://link.aps.org/doi/10.1103/PhysRevLett.113.236803}
}

@article{topo6,
doi = {10.1088/1367-2630/17/12/125014},
url = {https://dx.doi.org/10.1088/1367-2630/17/12/125014},
year = {2015},
month = {dec},
publisher = {IOP Publishing},
volume = {17},
number = {12},
pages = {125014},
author = {Nathan, Frederik and Rudner, Mark S},
title = {Topological singularities and the general classification of {F}loquet–{B}loch systems},
journal = {New Journal of Physics}
}

@article{topo7,
  title = {Signatures and conditions for phase band crossings in periodically driven integrable systems},
  author = {Mukherjee, Bhaskar and Sen, Arnab and Sen, Diptiman and Sengupta, K.},
  journal = {Phys. Rev. B},
  volume = {94},
  issue = {15},
  pages = {155122},
  numpages = {12},
  year = {2016},
  month = {Oct},
  publisher = {American Physical Society},
  doi = {10.1103/PhysRevB.94.155122},
  url = {https://link.aps.org/doi/10.1103/PhysRevB.94.155122}
}

@article{topo8,
  title = {Low-frequency phase diagram of irradiated graphene and a periodically driven spin-$\frac{1}{2}$ {XY} chain},
  author = {Mukherjee, Bhaskar and Mohan, Priyanka and Sen, Diptiman and Sengupta, K.},
  journal = {Phys. Rev. B},
  volume = {97},
  issue = {20},
  pages = {205415},
  numpages = {23},
  year = {2018},
  month = {May},
  publisher = {American Physical Society},
  doi = {10.1103/PhysRevB.97.205415},
  url = {https://link.aps.org/doi/10.1103/PhysRevB.97.205415}
}

@article{topo9,
  title = {{F}loquet topological transition by unpolarized light},
  volume = {98},
  ISSN = {2469-9969},
  url = {http://dx.doi.org/10.1103/PhysRevB.98.235112},
  DOI = {10.1103/physrevb.98.235112},
  number = {23},
  pages={235112},
  journal = {Phys. Rev. B},
  publisher = {American Physical Society (APS)},
  author = {Mukherjee,  Bhaskar},
  year = {2018},
  month = dec 
}

@article{scar1,
  title = {Dynamical Scar States in Driven Fracton Systems},
  author = {Pai, Shriya and Pretko, Michael},
  journal = {Phys. Rev. Lett.},
  volume = {123},
  issue = {13},
  pages = {136401},
  numpages = {5},
  year = {2019},
  month = {Sep},
  publisher = {American Physical Society},
  doi = {10.1103/PhysRevLett.123.136401},
  url = {https://link.aps.org/doi/10.1103/PhysRevLett.123.136401}
}

@article{scar2,
  title = {Collapse and revival of quantum many-body scars via {F}loquet engineering},
  author = {Mukherjee, Bhaskar and Nandy, Sourav and Sen, Arnab and Sen, Diptiman and Sengupta, K.},
  journal = {Phys. Rev. B},
  volume = {101},
  issue = {24},
  pages = {245107},
  numpages = {12},
  year = {2020},
  month = {Jun},
  publisher = {American Physical Society},
  doi = {10.1103/PhysRevB.101.245107},
  url = {https://link.aps.org/doi/10.1103/PhysRevB.101.245107}
}

@article{scar3,
  title = {Exact {F}loquet quantum many-body scars under {R}ydberg blockade},
  author = {Mizuta, Kaoru and Takasan, Kazuaki and Kawakami, Norio},
  journal = {Phys. Rev. Res.},
  volume = {2},
  issue = {3},
  pages = {033284},
  numpages = {13},
  year = {2020},
  month = {Aug},
  publisher = {American Physical Society},
  doi = {10.1103/PhysRevResearch.2.033284},
  url = {https://link.aps.org/doi/10.1103/PhysRevResearch.2.033284}
}

@article{scar4,
  title = {Many-body scar state intrinsic to periodically driven system},
  author = {Sugiura, Sho and Kuwahara, Tomotaka and Saito, Keiji},
  journal = {Phys. Rev. Res.},
  volume = {3},
  issue = {1},
  pages = {L012010},
  numpages = {6},
  year = {2021},
  month = {Feb},
  publisher = {American Physical Society},
  doi = {10.1103/PhysRevResearch.3.L012010},
  url = {https://link.aps.org/doi/10.1103/PhysRevResearch.3.L012010}
}

@article{scar5,
  title = {Dynamics of the vacuum state in a periodically driven {R}ydberg chain},
  author = {Mukherjee, Bhaskar and Sen, Arnab and Sen, Diptiman and Sengupta, K.},
  journal = {Phys. Rev. B},
  volume = {102},
  issue = {7},
  pages = {075123},
  numpages = {16},
  year = {2020},
  month = {Aug},
  publisher = {American Physical Society},
  doi = {10.1103/PhysRevB.102.075123},
  url = {https://link.aps.org/doi/10.1103/PhysRevB.102.075123}
}

@article{scar6,
  title = {Discrete Time-Crystalline Order Enabled by Quantum Many-Body Scars: Entanglement Steering via Periodic Driving},
  author = {Maskara, N. and Michailidis, A. A. and Ho, W. W. and Bluvstein, D. and Choi, S. and Lukin, M. D. and Serbyn, M.},
  journal = {Phys. Rev. Lett.},
  volume = {127},
  issue = {9},
  pages = {090602},
  numpages = {7},
  year = {2021},
  month = {Aug},
  publisher = {American Physical Society},
  doi = {10.1103/PhysRevLett.127.090602},
  url = {https://link.aps.org/doi/10.1103/PhysRevLett.127.090602}
}

@article{scar7,
  title = {Driving quantum many-body scars in the PXP model},
  volume = {106},
  ISSN = {2469-9969},
  url = {http://dx.doi.org/10.1103/PhysRevB.106.104302},
  DOI = {10.1103/physrevb.106.104302},
  pages={104302},
  number = {10},
  journal = {Phys. Rev. B},
  publisher = {American Physical Society (APS)},
  author = {Hudomal,  Ana and Desaules,  Jean-Yves and Mukherjee,  Bhaskar and Su,  Guo-Xian and Halimeh,  Jad C. and Papić,  Zlatko},
  year = {2022},
  month = sep 
}

@article{scar8,
  title = {Discrete Time Crystals Enforced by {F}loquet-{B}loch Scars},
  author = {Huang, Biao and Leung, Tsz-Him and Stamper-Kurn, Dan M. and Liu, W. Vincent},
  journal = {Phys. Rev. Lett.},
  volume = {129},
  issue = {13},
  pages = {133001},
  numpages = {7},
  year = {2022},
  month = {Sep},
  publisher = {American Physical Society},
  doi = {10.1103/PhysRevLett.129.133001},
  url = {https://link.aps.org/doi/10.1103/PhysRevLett.129.133001}
}

@article{hsf1,
  title = {Prethermal Fragmentation in a Periodically Driven Fermionic Chain},
  author = {Ghosh, Somsubhra and Paul, Indranil and Sengupta, K.},
  journal = {Phys. Rev. Lett.},
  volume = {130},
  issue = {12},
  pages = {120401},
  numpages = {6},
  year = {2023},
  month = {Mar},
  publisher = {American Physical Society},
  doi = {10.1103/PhysRevLett.130.120401},
  url = {https://link.aps.org/doi/10.1103/PhysRevLett.130.120401}
}

@article{hsf2,
  title = {Signatures of fragmentation for periodically driven fermions},
  author = {Ghosh, Somsubhra and Paul, Indranil and Sengupta, K.},
  journal = {Phys. Rev. B},
  volume = {109},
  issue = {21},
  pages = {214304},
  numpages = {14},
  year = {2024},
  month = {Jun},
  publisher = {American Physical Society},
  doi = {10.1103/PhysRevB.109.214304},
  url = {https://link.aps.org/doi/10.1103/PhysRevB.109.214304}
}

@article{hsf3,
  title = {Hilbert space fragmentation and exact scars of generalized {F}redkin spin chains},
  volume = {103},
  ISSN = {2469-9969},
  url = {http://dx.doi.org/10.1103/PhysRevB.103.L220304},
  DOI = {10.1103/physrevb.103.l220304},
  number = {22},
  pages={L220304},
  journal = {Phys. Rev. B},
  publisher = {American Physical Society (APS)},
  author = {Langlett,  Christopher M. and Xu,  Shenglong},
  year = {2021},
  month = jun 
}

@article{hsf4,
  title = {Floquet engineering of {H}ilbert space fragmentation in {S}tark lattices},
  volume = {109},
  ISSN = {2469-9969},
  url = {http://dx.doi.org/10.1103/PhysRevB.109.184313},
  DOI = {10.1103/physrevb.109.184313},
  number = {18},
  pages={184313},
  journal = {Phys. Rev. B},
  publisher = {American Physical Society (APS)},
  author = {Zhang,  Li and Ke,  Yongguan and Lin,  Ling and Lee,  Chaohong},
  year = {2024},
  month = may 
}

@misc{hsf5,
      title={Destructive Interference induced constraints in {F}loquet systems}, 
      author={Somsubhra Ghosh and Indranil Paul and K. Sengupta and Lev Vidmar},
      year={2025},
      eprint={2508.18368},
      archivePrefix={arXiv},
      primaryClass={cond-mat.str-el},
      url={https://arxiv.org/abs/2508.18368}, 
}

@article{hsf6,
  title = {Signatures of fragmentation for periodically driven fermions},
  author = {Ghosh, Somsubhra and Paul, Indranil and Sengupta, K.},
  journal = {Phys. Rev. B},
  volume = {109},
  issue = {21},
  pages = {214304},
  numpages = {14},
  year = {2024},
  month = {Jun},
  publisher = {American Physical Society},
  doi = {10.1103/PhysRevB.109.214304},
  url = {https://link.aps.org/doi/10.1103/PhysRevB.109.214304}
}

@misc{meissner1,
      title={Floquet realization of prethermal 
      {M}eissner phase in a two-leg flux ladder}, 
      author={Biswajit Paul and Tapan Mishra and K. Sengupta},
      year={2025},
      eprint={2504.11017},
      archivePrefix={arXiv},
      primaryClass={cond-mat.quant-gas},
      url={https://arxiv.org/abs/2504.11017}, 
}

@article{berry1,
doi = {10.1088/1751-8113/42/36/365303},
url = {https://doi.org/10.1088/1751-8113/42/36/365303},
year = {2009},
month = {aug},
publisher = {},
volume = {42},
number = {36},
pages = {365303},
author = {Berry, M V},
title = {Transitionless quantum driving},
journal = {Journal of Physics A}

}

@Article{rev18,
author={del Campo, A.
and Sengupta, K.},
title={Controlling quantum critical dynamics of isolated systems},
journal={The European Physical Journal Special Topics},
year={2015},
month={Feb},
day={01},
volume={224},
number={1},
pages={189-203},
issn={1951-6401},
doi={10.1140/epjst/e2015-02350-4},
url={https://doi.org/10.1140/epjst/e2015-02350-4}
}

@article{tista1,
doi = {10.1088/1367-2630/adfd07},
url = {https://doi.org/10.1088/1367-2630/adfd07},
year = {2025},
month = {aug},
publisher = {IOP Publishing},
volume = {27},
number = {8},
pages = {084506},
author = {Banerjee, Tista and Choudhury, Sayan and Sengupta, K},
title = {Exact {F}loquet flat band and heating suppression via two-tone drive protocols},
journal = {New Journal of Physics}
}

@misc{sm1,
      title={The {K}itaev-{AKLT} model}, 
      author={Alwyn Jose Raja and R. Ganesh},
      year={2025},
      eprint={2510.12880},
      archivePrefix={arXiv},
      primaryClass={cond-mat.str-el},
      url={https://arxiv.org/abs/2510.12880}, 
}

@article{sm2,
  title = {Pronounced quantum many-body scars in the one-dimensional spin-1 {K}itaev model},
  author = {Mohapatra, Sashikanta and Balram, Ajit C.},
  journal = {Phys. Rev. B},
  volume = {107},
  issue = {23},
  pages = {235121},
  numpages = {8},
  year = {2023},
  month = {Jun},
  publisher = {American Physical Society},
  doi = {10.1103/PhysRevB.107.235121},
  url = {https://link.aps.org/doi/10.1103/PhysRevB.107.235121}
}

@article{dsen1,
  title = {Spin-1 {K}itaev model in one dimension},
  author = {Sen, Diptiman and Shankar, R. and Dhar, Deepak and Ramola, Kabir},
  journal = {Phys. Rev. B},
  volume = {82},
  issue = {19},
  pages = {195435},
  numpages = {11},
  year = {2010},
  month = {Nov},
  publisher = {American Physical Society},
  doi = {10.1103/PhysRevB.82.195435},
  url = {https://link.aps.org/doi/10.1103/PhysRevB.82.195435}
}

@article{fss1,
  title = {Competing density-wave orders in a one-dimensional hard-boson model},
  author = {Fendley, Paul and Sengupta, K. and Sachdev, Subir},
  journal = {Phys. Rev. B},
  volume = {69},
  issue = {7},
  pages = {075106},
  numpages = {15},
  year = {2004},
  month = {Feb},
  publisher = {American Physical Society},
  doi = {10.1103/PhysRevB.69.075106},
  url = {https://link.aps.org/doi/10.1103/PhysRevB.69.075106}
}

@article{page1,
  title = {Average entropy of a subsystem},
  author = {Page, Don N.},
  journal = {Phys. Rev. Lett.},
  volume = {71},
  issue = {9},
  pages = {1291--1294},
  numpages = {0},
  year = {1993},
  month = {Aug},
  publisher = {American Physical Society},
  doi = {10.1103/PhysRevLett.71.1291},
  url = {https://link.aps.org/doi/10.1103/PhysRevLett.71.1291}
}

@article{saito1,
title = {Floquet–{M}agnus theory and generic transient dynamics in periodically driven many-body quantum systems},
journal = {Annals of Physics},
volume = {367},
pages = {96-124},
year = {2016},
issn = {0003-4916},
doi = {https://doi.org/10.1016/j.aop.2016.01.012},
url = {https://www.sciencedirect.com/science/article/pii/S0003491616000142},
author = {Tomotaka Kuwahara and Takashi Mori and Keiji Saito}
}

@article{mori1,
  title = {Rigorous Bound on Energy Absorption and Generic Relaxation in Periodically Driven Quantum Systems},
  author = {Mori, Takashi and Kuwahara, Tomotaka and Saito, Keiji},
  journal = {Phys. Rev. Lett.},
  volume = {116},
  issue = {12},
  pages = {120401},
  numpages = {5},
  year = {2016},
  month = {Mar},
  publisher = {American Physical Society},
  doi = {10.1103/PhysRevLett.116.120401},
  url = {https://link.aps.org/doi/10.1103/PhysRevLett.116.120401}
}

@Article{da1,
author={Abanin, Dmitry
and De Roeck, Wojciech
and Ho, Wen Wei
and Huveneers, Fran{\c{c}}ois},
title={A Rigorous Theory of Many-Body Prethermalization for Periodically Driven and Closed Quantum Systems},
journal={Communications in Mathematical Physics},
year={2017},
month={Sep},
day={01},
volume={354},
number={3},
pages={809-827},
issn={1432-0916},
doi={10.1007/s00220-017-2930-x},
url={https://doi.org/10.1007/s00220-017-2930-x}
}

@article{da2,
  title = {Effective Hamiltonians, prethermalization, and slow energy absorption in periodically driven many-body systems},
  author = {Abanin, Dmitry A. and De Roeck, Wojciech and Ho, Wen Wei and Huveneers, Fran\ifmmode \mbox{\c{c}}\else \c{c}\fi{}ois},
  journal = {Phys. Rev. B},
  volume = {95},
  issue = {1},
  pages = {014112},
  numpages = {8},
  year = {2017},
  month = {Jan},
  publisher = {American Physical Society},
  doi = {10.1103/PhysRevB.95.014112},
  url = {https://link.aps.org/doi/10.1103/PhysRevB.95.014112}
}

@article{dc1,
  title = {Floquet Thermalization via Instantons near Dynamical Freezing},
  author = {Mukherjee, Rohit and Guo, Haoyu and Chowdhury, Debanjan},
  journal = {Phys. Rev. X},
  volume = {16},
  issue = {1},
  pages = {011041},
  numpages = {26},
  year = {2026},
  month = {Feb},
  publisher = {American Physical Society},
  doi = {10.1103/4w5w-57my},
  url = {https://link.aps.org/doi/10.1103/4w5w-57my}
}

@article{vk1,
  title = {Prethermal Stability of Eigenstates under High Frequency Floquet Driving},
  author = {O'Dea, Nicholas and Burnell, Fiona and Chandran, Anushya and Khemani, Vedika},
  journal = {Phys. Rev. Lett.},
  volume = {132},
  issue = {10},
  pages = {100401},
  numpages = {6},
  year = {2024},
  month = {Mar},
  publisher = {American Physical Society},
  doi = {10.1103/PhysRevLett.132.100401},
  url = {https://link.aps.org/doi/10.1103/PhysRevLett.132.100401}
}

@article{vk2,
  title = {Phenomenology of the Prethermal Many-Body Localized Regime},
  author = {Long, David M. and Crowley, Philip J. D. and Khemani, Vedika and Chandran, Anushya},
  journal = {Phys. Rev. Lett.},
  volume = {131},
  issue = {10},
  pages = {106301},
  numpages = {7},
  year = {2023},
  month = {Sep},
  publisher = {American Physical Society},
  doi = {10.1103/PhysRevLett.131.106301},
  url = {https://link.aps.org/doi/10.1103/PhysRevLett.131.106301}
}

@article{aditya2024,
  title = {Subspace-restricted thermalization in a correlated-hopping model with strong {H}ilbert space fragmentation characterized by irreducible strings},
  author = {Aditya, Sreemayee and Dhar, Deepak and Sen, Diptiman},
  journal = {Phys. Rev. B},
  volume = {110},
  issue = {4},
  pages = {045418},
  numpages = {19},
  year = {2024},
  month = {Jul},
  publisher = {American Physical Society},
  doi = {10.1103/PhysRevB.110.045418},
  url = {https://link.aps.org/doi/10.1103/PhysRevB.110.045418}
}

@article{ganguli2025,
  title = {Aspects of {H}ilbert space fragmentation in the quantum {E}ast model: Fragmentation, subspace-restricted quantum scars, and effects of density-density interactions},
  author = {Ganguli, Maitri and Aditya, Sreemayee and Sen, Diptiman},
  journal = {Phys. Rev. B},
  volume = {111},
  issue = {4},
  pages = {045411},
  numpages = {25},
  year = {2025},
  month = {Jan},
  publisher = {American Physical Society},
  doi = {10.1103/PhysRevB.111.045411},
  url = {https://link.aps.org/doi/10.1103/PhysRevB.111.045411}
}

@article{moudgalya2022,
doi = {10.1088/1361-6633/ac73a0},
url = {https://doi.org/10.1088/1361-6633/ac73a0},
year = {2022},
month = {jul},
publisher = {IOP Publishing},
volume = {85},
number = {8},
pages = {086501},
author = {Moudgalya, Sanjay and Bernevig, B Andrei and Regnault, Nicolas},
title = {Quantum many-body scars and {H}ilbert space fragmentation: a review of exact results},
journal = {Reports on Progress in Physics},
}

@article{sala2020,
  title = {Ergodicity Breaking Arising from 
  {H}ilbert Space Fragmentation in Dipole-Conserving Hamiltonians},
  author = {Sala, Pablo and Rakovszky, Tibor and Verresen, Ruben and Knap, Michael and Pollmann, Frank},
  journal = {Phys. Rev. X},
  volume = {10},
  issue = {1},
  pages = {011047},
  numpages = {19},
  year = {2020},
  month = {Feb},
  publisher = {American Physical Society},
  doi = {10.1103/PhysRevX.10.011047},
  url = {https://link.aps.org/doi/10.1103/PhysRevX.10.011047}
}

@article{khemani2020,
  title = {Localization from {H}ilbert space shattering: From theory to physical realizations},
  author = {Khemani, Vedika and Hermele, Michael and Nandkishore, Rahul},
  journal = {Phys. Rev. B},
  volume = {101},
  issue = {17},
  pages = {174204},
  numpages = {17},
  year = {2020},
  month = {May},
  publisher = {American Physical Society},
  doi = {10.1103/PhysRevB.101.174204},
  url = {https://link.aps.org/doi/10.1103/PhysRevB.101.174204}
}

\end{document}